\documentclass[10pt,letterpaper]{article}
\usepackage[top=0.85in,left=2.75in,footskip=0.75in]{geometry}
\usepackage{amsmath,amssymb}
\usepackage{changepage}
\usepackage{textcomp,marvosym}
\usepackage{cite}
\usepackage{nameref,hyperref}
\hypersetup{hidelinks}
\usepackage[nopatch=eqnum]{microtype}
\ifdefined\pdftexversion
\DisableLigatures[f]{encoding = *, family = * }
\fi
\usepackage[table]{xcolor}
\usepackage{array}
\newcolumntype{+}{!{\vrule width 2pt}}
\newlength\savedwidth

\newcommand{\sitablefont}{\scriptsize\setlength{\tabcolsep}{4pt}\renewcommand{\arraystretch}{1.05}}
\raggedright
\usepackage[aboveskip=1pt,labelfont=bf,labelsep=period,justification=raggedright,singlelinecheck=off]{caption}

\makeatletter
\renewcommand{\@biblabel}[1]{\quad#1.}
\makeatother
\usepackage{lastpage,fancyhdr,graphicx}
\ifdefined\pdftexversion
\usepackage{epstopdf}
\fi
\fancyheadoffset[L]{2.25in}
\fancyfootoffset[L]{2.25in}
\begin{document}
\vspace*{0.2in}

\begin{flushleft}
{\Large
\textbf{Conflict, mobility and fragmentation in the African interurban network}
}
\newline
\\
Andrew Renninger\textsuperscript{1,*},
Valentina Marin\textsuperscript{2},
Carmen Cabrera\textsuperscript{3},
Rafael Prieto-Curiel\textsuperscript{4}
\\
\bigskip
\textbf{1} Central European University, Vienna, Austria 
\\
\textbf{2} University College London, London, United Kingdom 
\\
\textbf{3} University of Liverpool, Liverpool, United Kingdom 
\\
\textbf{4} Complexity Science Hub, Vienna, Austria
\\
\bigskip
* Corresponding author: renningera@ceu.edu
\end{flushleft}

\section*{Abstract}

Roads make regional integration possible, but they also concentrate risk. When violence reaches a corridor that carries regional movement, the consequences can extend beyond the attacked place. We study this problem in Africa by combining interurban road data, georeferenced conflict events, and bilateral migration flows. Using a Hawkes-style self-exciting memory kernel, we estimate accumulated conflict intensity near roads from 2019 to 2022, and we route modelled movement between cities along the road network. As an empirical check on these modelled corridor flows, we show that country pairs connected by routes with higher accumulated conflict have lower bilateral migration. The association remains negative across alternative assumptions about spatial decay and is strongest for paths that cross regions, consistent with a corridor-deterrence channel rather than only displacement from violent places. We then use cascade and percolation simulations to examine how local disruptions propagate through the road network. Most simulated disruptions isolate relatively small populations, but 18.6\% isolate more than one million people, and the largest isolates 13.3 million. Removing roads in order of conflict intensity reveals a tipping point, at which the giant connected component falls from 85\% to 39\% of the network. Our results identify the roads where violence most threatens regional integration and show how past conflict, mobility, and network structure combine to create persistent corridor vulnerability.

\section*{Author summary}

Roads connect cities to markets, services, and each other, but the same corridors can become points of vulnerability when violence makes travel dangerous or impossible. We study this problem across Africa by combining records of conflict events, a map of major interurban roads, city populations, and migration flows between countries. We use the migration data to check whether the corridors that our model identifies as important also matter for observed movement, and whether conflict along those corridors is associated with lower movement. Conflict along road corridors is associated with lower bilateral migration, even after accounting for baseline movement predicted from population, distance, and intervening opportunities. We then simulate how local disruptions can spread through nearby roads or accumulate across the network. Most disruptions affect small populations, but a small number can isolate millions of urban residents from the main connected road system. Our results show that intense conflict threatens corridors that organize movement and can weaken regional integration by exposing fragile parts of the continental transport network.

\clearpage
\newgeometry{top=0.85in,left=1in,right=1in,footskip=0.75in}

\section*{Introduction}
Regional integration depends on transport networks that keep cities and markets connected across long distances. With the African Continental Free Trade Area in force, the question is not only whether trade barriers fall on paper, but whether firms and consumers can reach regional and international markets in practice \cite{worldbank2012defragmenting, unctad2019madeinafrica, maliszewska2020afcfta, afdb2019aeo}. Transport costs, border procedures, road quality, logistics, and other frictions still shape that access \cite{maliszewska2020afcfta, unctad2019madeinafrica, afdb2019aeo, worldbank2012defragmenting, teravaninthorn2009transport}. Yet violent conflict across parts of the continent threatens the roads on which regional integration depends \cite{worldbank2011wdr, fang2020consequences, raleigh2010introducing, chathamTerrorism, martin2008maketrade, glick2010collateraldamage, assem2023tradeconflicts}. Understanding how conflict disrupts road corridors is therefore central to assessing when regional integration can generate broad economic gains \cite{mueller2013costsconflict, maliszewska2020afcfta}.

The consequences of conflict are not limited to the places where violence occurs. Roads link cities, countries, and markets, so violence on one segment can lengthen journeys, redirect movement, or sever access for places that are not directly exposed to conflict. These knock-on effects matter most where there are few alternative routes. Landlocked countries are among the least integrated in Africa \cite{limao2001infrastructure, worldbank2012defragmenting, teravaninthorn2009transport}, and infrastructure constraints can account for as much as 60\% of transport costs in inland economies, compared with 40\% in coastal ones \cite{limao2001infrastructure}. For inland economies and border regions that depend on a small number of roads to reach regional and international markets, conflict may therefore create costs far from the event itself. Yet many of the places that depend on these corridors, including Chad, Niger, and Burkina Faso, also sit near zones where conflict is high or rising \cite{raleigh2010introducing, chathamTerrorism, prieto2024geographical, fang2020consequences}.

Existing research gives pieces of the answer. Studies of regional integration show that transport costs, border delays, and logistics constraints limit exchange across borders \cite{maliszewska2020afcfta, lebrand2021corridors, teravaninthorn2009transport}. Studies of conflict show that violence depresses trade, disrupts mobility, and imposes costs on neighboring economies \cite{martin2008maketrade, glick2010collateraldamage, assem2023tradeconflicts}. Georeferenced conflict and infrastructure data now make it possible to study violence and transport systems at fine spatial scales \cite{raleigh2010introducing, prieto2022detecting, prieto2022constructing}. What remains harder to see is the joint margin: which road corridors are both important for movement and exposed to conflict, whether mobility falls when routes between countries accumulate violence, and where disruptions to exposed roads would most threaten regional connectivity.

We combine three sources of evidence on the African interurban road network. First, we use georeferenced conflict events from the Armed Conflict Location \& Event Data project (ACLED) to estimate the spatial and temporal intensity of violence near roads \cite{raleigh2010introducing}. We use a Hawkes-style self-exciting memory kernel to construct conflict intensity measures for network nodes and road segments, rather than estimating a full generative Hawkes point process. Second, we combine Africapolis city data with a connected network of major roads constructed from OpenStreetMap to model expected movement between cities \cite{Africapolis, OpenStreetMap, prieto2022detecting, prieto2022constructing}. We route those interactions over shortest paths, giving projected movement values for roads and cities. Third, we compare conflict exposure along road paths with observed bilateral migration flows from 2019 to 2022. This lets us test whether conflict on transport corridors is associated with reduced mobility between countries, rather than treating vulnerability as a simulated property of the network alone.

The analysis yields three results. First, projected movement and conflict exposure are unevenly distributed across the road network and do not always coincide. Some roads are structurally important but remain comparatively peaceful, while others combine strong dependence with persistent conflict. Second, observed migration grounds the corridor interpretation: country pairs connected by road paths with higher accumulated conflict intensity have lower bilateral migration from 2019 to 2022. This association remains negative across alternative assumptions about spatial decay and is strongest for paths that cross regions, consistent with conflict making regional corridors harder to use rather than only displacing people from the immediate location of violence. Third, cascade and percolation simulations show that local disruptions can have highly uneven network consequences. Most simulated cascades isolate relatively small populations, but 18.6\% isolate more than one million people, and the largest isolates 13.3 million. When roads are removed in order of conflict intensity, the network also exhibits a fragmentation threshold at which the giant connected component falls from 85\% to 39\% of the network. The resulting geography of vulnerability is concentrated in parts of West and East Africa, including corridors affected by groups such as Boko Haram and Al-Shabaab.

\section*{Context}
\subsection*{Violence against civilians in Africa is increasing and concentrated} 
ACLED records the location, date, actors, fatalities, and type of reported political violence and protest events worldwide, primarily from local media sources \cite{raleigh2010introducing}. That makes it especially useful for connecting violence to roads, cities, and borders, even though it remains a record of reported events rather than a full census of insecurity. Between 2010 and 2022, the ACLED extract used here records 159,057 conflict events in Africa, with 405,851 reported fatalities and 60,484 events classified as violence against civilians. In 2021 alone, 23.5 events per day were classified as violence against civilians, causing 43.6 fatalities per day; by 2022, those figures had risen to 24.7 events and 47.7 fatalities per day. The event frequency was more than five times higher than ten years earlier.

Three features of this violence matter for a network analysis. First, casualties are highly concentrated \cite{ZipfTerrorism, guo2019common}. Of the roughly 8,600 events classified as violence against civilians in 2021, the top 5\% accounted for 52\% of casualties; roughly one event per day accounted for half of civilian deaths. Second, violence is often urban \cite{radil2022urban} and, in Africa, more lethal in smaller and more isolated cities \cite{prieto2024geographical}. Third, violence is uneven and persistent. Mogadishu had nearly 100 casualties from violence against civilians in 2021 and more than 100 in many earlier years, while Abuja, with a similar population, recorded one casualty in 2021 and remained at low levels for more than a decade \cite{raleigh2010introducing, prieto2024geographical, ViolenceNorthWestAfrica}. The relevant risk surface is therefore not only where violence occurred once, but where it recurs.

Conflict also damages the economic systems around violence. Countries experiencing intense civil conflict grow 2.5 percentage points more slowly on average \cite{fang2020consequences}, and civil wars reduce trade flows by as much as 40\%, with larger effects in more intense conflicts \cite{martin2008maketrade, glick2010collateraldamage, assem2023tradeconflicts}. Over a protracted conflict, losses compound in ways that matter for infrastructure: governments defer maintenance, forgo new projects, and leave roads to be destroyed or neglected \cite{worldbank2011wdr, fang2020consequences, gates2010consequences}. By one estimate, civil war can set infrastructure coverage back by a decade or more \cite{worldbank2011wdr, gates2010consequences}.

The spatial position of violence matters as much as its intensity. In North and West Africa, about 70\% of recorded violent events and 65\% of fatalities occur within 1\,km of a road; in West Africa, the share near roads has fallen as conflict has spread into more remote spaces \cite{OECDSWACRoadsConflicts2025, raleigh2010introducing}. Roads make markets, services, and movement across borders reachable, but they are also visible, strategic, and vulnerable to groups seeking control over territory and mobility \cite{chathamTerrorism, prieto2020uncovering, OECDSWACRoadsConflicts2025}.

Conflict has memory. A recent attack can change the risk of subsequent attacks, but that effect should attenuate over time rather than hold constant; a Hawkes-style memory kernel gives us a way to represent that memory without treating all past violence as equivalent \cite{hawkes1971spectra, MohlerExcitingPointProcess, laub2015hawkes}. The same logic appears in earthquakes, communication, and other self-exciting processes: a shock raises the likelihood of another event nearby in time, then its influence fades unless new events arrive \cite{chowdhary2023temporal, porter2012self, yuan2019fast, telesca2006global, clauset2013estimating}. In addition to frequency, the model also needs amplitude. An event with many deaths should not enter the risk surface in the same way as an event with none. We therefore use fatalities to weight event severity while still allowing nonfatal events to contribute to the intensity surface \cite{ZipfTerrorism}.

\subsection*{Road networks are vulnerable} 
Roads do different jobs: some are local connectors, while others link cities to ports, markets, capitals, and border crossings. African geography and institutions sharpen that distinction: low road density, high freight prices, landlockedness, border delays, checkpoints, and informal payments all make corridor performance a condition for market integration \cite{worldbank2012defragmenting, teravaninthorn2009transport, walther2020mapping, BensassiJarreau2019}. Research on transport costs finds that poor infrastructure explains a large share of the cost penalty facing coastal areas and an even larger share in inland countries \cite{limao2001infrastructure, storeygard2016farther}. Counterfactual modeling in West Africa makes the same point: road improvements generate welfare gains, but those gains are larger and more distributed when border delays fall at the same time \cite{lebrand2021corridors, walther2020mapping}. Connectivity depends on more than links on a map. It depends on the time, cost, reliability, and political control of the corridor between them.

The loss of a minor road may not affect many journeys, but losing a major corridor can disconnect regions or lengthen journey times substantially. Network disruption and percolation methods formalize that difference by removing nodes or roads, measuring how damage propagates, and asking when many local removals become a failure of the connected system \cite{crucitti2004error, cascadingPorta2004, newman2003structure, barabasi2016network}. Similar threshold behavior appears in infrastructure, disease, brain, and urban systems, where the largest connected component gives a concise measure of whether the system still holds together \cite{Callaway2000, newman2002spread, Gallos2011, Gallos2012, Fluschnik2016, Piovani2017, Behnisch2019, marin2023scalar}.

Prior work on spatial networks measures disruption through longer paths, lost accessibility, changed centrality, rerouted flows, and the number of places that remain reachable after a shock \cite{gil2008flood, maureira2017everyday, PaoloMasucci2016, berche2009resilience, Derrible2010, pagani2019resilience, newman2003structure}. These effects depend on the type of incident and on the position of the disrupted element within the system \cite{Holme2002}. Conflict makes that positional problem sharper because it can raise transport costs, close borders, force rerouting, and spread losses to neighboring economies through shared transit routes \cite{assem2023tradeconflicts, harari2018conflict, glick2010collateraldamage}. Many African transport systems also lack redundancy, so a single road or port closure can sever links that have few substitutes \cite{UNTransportAfrica09, teravaninthorn2009transport, ganin2017resilience}. The same road failure can be minor on a redundant corridor and severe where few alternatives exist.

City position also shapes economic opportunity. Cities near large urban regions can borrow economies of agglomeration \cite{meijers2016borrowing}, and economic activity often responds to geographic centrality \cite{StreetNetw}. Proximity alone, however, does not capture accessibility. Geographic barriers, border crossings, travel times, and the road network itself determine which cities can exchange people and goods \cite{barbosa2018human, walther2020mapping, storeygard2016farther}. 

Network importance can mean many things, from simple degree to centrality measures that account for position relative to other nodes \cite{buechel2013dynamics, singh2020node, freeman1978centrality}. We are concerned not with how central a city or road is in the abstract, but how much expected movement depends on it. Work on road network vulnerability measures link importance by the travel costs, accessibility losses, rerouted trips, or delays created by disruption \cite{berdica2002introduction, jenelius2006importance, taylor2006accessibility, scott2006network, jenelius2009network, mattsson2015vulnerability, li2020critical}. We adopt the same logic by estimating movement between cities with a radiation model and assigning those flows to the shortest road paths that carry them. Radiation models offer a parsimonious way to estimate mobility and migration from population and intervening opportunities, and they require no fitted parameters where detailed calibration data are limited \cite{simini2012universal, lenormand2016systematic}. If conflict disrupts one of these routes, the consequences for intercity travel can extend beyond the immediate location of violence because checkpoints, blockades, insecurity, and contested control of roads change which places remain reachable \cite{schouten2022roadblock, agbiboa2022mobility, ali2015infrastructure, oecdSWAC2025roadsConflicts}. The integration question is which connections keep movement possible when violence makes parts of the system unusable \cite{mullercrepon2021roads, oecdSWAC2025roadsConflicts}. A route through a city or road can matter because it carries many trips, because there are few substitutes, or because both are true; the cascade and percolation exercises ask how many people lose access to the main road system when such places fail \cite{ArcautePercolation16, baggag2018resilience, schuster2023stress}.

\section*{Materials and methods}
\subsection*{Measuring conflict intensity}
The effects of conflict depend on both frequency and amplitude---a city with a rapid succession of attacks is different from a city with occasional attacks over many years; an attack in which many people die is different from an attack with no reported fatalities---and our models need to account for that. We construct a Hawkes-style self-exciting exposure index: each event raises local intensity, and that contribution decays until new events arrive. We use the Hawkes memory kernel to measure accumulated exposure, not to estimate a full generative point process for conflict. For network node $i$ at time $t$, conflict intensity is

\begin{equation} 
\label{hawkes}
\mu_i(t) = \sum_{s: t \geq t_s} I_i(s) \, \kappa_s \, \exp\!\big(-\phi (t-t_s)\big),
\end{equation}

\noindent
where $I_i(s) = 1$ if event $s$ occurred at node $i$ and zero otherwise, $\kappa_s$ weights the event, and $\phi$ governs the rate at which past events fade. The half-life of an event's contribution is $\ln(2)/\phi$.

We weight each event by $\kappa_s = 1 + f_s$, where $f_s$ is the reported number of fatalities. The additive shift keeps events without recorded fatalities from dropping out of the intensity calculation; without it, only lethal violence would register. For the migration analysis, we assign conflict events to the nearest node in the road network within 10\,km, allowing both cities and road junctions to contribute to corridor exposure \cite{laub2015hawkes}.

Road-level conflict intensity, $\mu_e$, is the edge-risk index used for the road maps, cascades, detours, and attack percolation. It accumulates conflict exposure near each road segment rather than at named cities only, using fatality-weighted event contributions and temporal decay on the same intensity scale. This distinction matters because $\mu_e$ describes road exposure, while the city table in the Supporting information reports city-level averages and peaks from the node surface.

The decay rate should reflect what the measure is being asked to explain. For prediction over the next two weeks, we choose $\phi$ by grid search over $[0.005, 0.55]$, selecting the value that maximizes the correlation between current intensity and observed fatalities at the same node over the following two weeks. This gives $\phi^\star = 0.055$, a half-life of roughly 13 days. We use this calibration for the cascade and percolation analyses. The migration analysis uses a slower main exposure, $\phi = 0.0005$ (half-life ${\sim}1{,}386$ days), because movement between countries may respond to a corridor's accumulated history rather than only its most recent events; the Hawkes calibration is shown in Supplementary Fig.~\ref{S2_Fig}, and the memory checks for migration are shown in Supplementary Fig.~\ref{S4_Fig}.

\subsection*{Estimating expected movement}
The road network tells us which cities could be connected, but not the degree to which these connections organize movement. We use a network radiation model because it gives more weight to large destinations while discounting places with many intervening opportunities along the route \cite{barbosa2018human}. This makes expected movement depend on the structure of the corridor as well as population and distance. For cities $i$ and $j$, expected movement is

\begin{equation}
\label{radiation}
R_{ij} \propto
\frac{P_i P_j}{(P_i + s_{ij})(P_i + P_j + s_{ij})},
\end{equation}

\noindent
where $P_i$ and $P_j$ are city populations and $s_{ij}$ is the population on the shortest road path between cities $i$ and $j$, excluding the endpoints. We cap paths at 3{,}500 km, approximately the road distance from Lagos to Dakar, a major economic corridor spanning the West African coast \cite{lebrand2021corridors}. The radiation baseline better preserves the rank ordering of country pairs with high or low migration in the observed data; the baseline comparison is shown in Supplementary Fig.~\ref{S4_Fig}, and the gravity calibration is shown in Supplementary Fig.~\ref{S5_Fig}.

The radiation model gives us expected movement between cities; the network question requires movement along roads. Let $\widetilde{R}_{ij}$ denote the normalized radiation flow for city pair $i,j$. We assign each $\widetilde{R}_{ij}$ to the shortest path between the cities and sum the flows that use each road $e$:

\begin{equation}
\label{edge_flow}
\nu_e = \sum_{i,j: e \in \pi(i,j)} \widetilde{R}_{ij},
\end{equation}

\noindent
where $\pi(i,j)$ is the shortest road path between the two cities. The resulting quantity, $\nu_e$, measures how much expected movement depends on each road. We compare radiation with a suite of calibrated gravity models in Supplementary Figs.~\ref{S4_Fig} and \ref{S5_Fig}. Radiation requires no fitted distance decay parameter and better preserves the rank ordering of observed migration corridors, while gravity provides the robustness baseline with lower RMSE.

\subsection*{Grounding corridor exposure in observed migration}
Observed migration enters the paper as a grounding check, not as the main outcome. We use it to check whether a road network model of spatial interaction places movement on plausible corridors, and whether conflict along those corridors is associated with lower observed movement. The migration data come from monthly flows between countries described by Chi et al., based on privacy protected Facebook records and adjusted for variation in platform use across countries \cite{chi2025measuring}. We aggregate the monthly estimates to directed country pairs by year from 2019 to 2022 and convert flows to shares of each origin's observed migration.

For each origin country $o$ and destination country $d$, we route all city pairs $i \in o$ and $j \in d$ through the road network and weight each path by the city flow $R_{ij}$ modelled by radiation. Corridor conflict is the weighted mean of conflict along the shortest path between those cities,

\begin{equation}
\label{path_conflict}
C_{od,y} =
\frac{\sum_{i \in o}\sum_{j \in d} R_{ij}\left(
|\pi(i,j)|^{-1}\sum_{k \in \pi(i,j)} \mu_{k,y}
\right)}
{\sum_{i \in o}\sum_{j \in d} R_{ij}},
\end{equation}

\noindent
where $\pi(i,j)$ is the shortest road path from city $i$ to city $j$, and $\mu_{k,y}$ is the Hawkes intensity at node $k$ in year $y$. The inner mean keeps longer paths between cities from being assigned more conflict simply because they pass through more nodes, while the outer weighted mean lets city pairs with larger predicted movement contribute more to the corridor exposure for the country pair.

We then estimate whether observed migration is lower on corridors with more conflict after accounting for the migration predicted by the radiation model:

\begin{equation}
\label{migration_regression}
\log_{10} M_{od,y}
= \alpha
+ \beta_1 \log_{10} \hat{M}_{od,y}
+ \beta_2 \log_{10}(C_{od,y} + 0.01)
+ \epsilon_{od,y}.
\end{equation}

\noindent
Here $M_{od,y}$ is the observed migration share, $\hat{M}_{od,y}$ is the share predicted by the radiation model, and $C_{od,y}$ is corridor conflict. The small constant keeps corridors with very low conflict in the log scale. The coefficient of interest is $\beta_2$: it asks whether migration is lower than expected when the weighted road corridors between two countries are more exposed to conflict. For our purposes, this is a grounding test for the modelled corridor flows rather than the paper's main claim. We estimate the model overall, by year, by origin region, and separately for routes within and across regions. Standard errors are clustered by directed country pair; in year-specific models each pair contributes one observation, so the clustered estimate reduces to pair-level heteroskedasticity-robust uncertainty.

\subsection*{Simulating local conflict cascades}
Local violence can disrupt more than the road where it occurs. The cascade model asks how much isolation can follow when nearby roads with high conflict intensity also become unusable. Each simulation starts by selecting a seed road with probability proportional to its conflict intensity:

\begin{equation}
\label{seed_probability}
\Pr(\mathrm{seed}=e) =
\frac{\mu_e}{\sum_h \mu_h}.
\end{equation}

\noindent
Nearby roads then fail with a probability that depends on their own conflict exposure and their Euclidean distance from the seed. For road $f$ given seed road $e$,

\begin{equation}
\label{cascade_probability}
\Pr(f \ \mathrm{fails} \mid e)
= 1 - \exp\!\left(-\frac{\mu_f}{d(e,f)^\alpha}\right),
\end{equation}

\noindent
where $d(e,f)$ is the Euclidean distance between road midpoints in kilometres and $\alpha=1.1$ is the spatial decay parameter. We floor $d(e,f)$ at 1\,km to avoid singularities for adjacent or overlapping segments, and $\mu_f$ is used on the edge-risk scale described above. The main power-law cascade has no hard maximum radius: every road has a nonzero probability of failing, but that probability decays quickly with distance and is effectively concentrated near the seed. We use Euclidean distance because conflict spillovers are not necessarily constrained to the road network; armed groups, displacement, information, and retaliation can move through geographic proximity as well as along roads. The value of $\alpha$ comes from the observed spatial decay of ACLED events, which follows a power law more closely than an exponential decay.

In each of 10,000 simulations, we remove the seed road and any roads that fail the Bernoulli draw in Eq.~\ref{cascade_probability}. We then rebuild the network and measure the number of failed roads, the number of isolated city nodes, the population in those isolated cities, and the dominant region affected. The network is reset after every simulation, so this model captures localized cascade risk rather than cumulative continental collapse.

This model assumes that violence propagates through both \emph{time} and \emph{space}. Supplementary Fig.~\ref{S6_Fig} tests this assumption using Knox and Local Knox tests: after date shuffling within country and year, violent events are still followed by excess nearby violence over the next 13 days and as far as 200\,km away. As a reporting-based robustness check, we also use 2024 West African news reports extracted from the GDELT Project that mention corridor, border-road, or border-closure disruption \cite{gdeltproject}. Peaks in city-level Hawkes intensity are followed by excess disruption mentions in other nearby cities, while same-city pairs do not account for the pattern (Supplementary Fig.~\ref{S10_Fig}).

\subsection*{Measuring network fragmentation}
A second question is where accumulated conflict risk begins to fragment the whole network. We answer this with attack percolation, removing roads above a threshold in conflict intensity and measuring the connected network that remains \cite{ArcautePercolation16}.

For a threshold $\tau$, the removed road set is

\begin{equation}
\label{percolation_threshold}
R(\tau) = \{e : \mu_e > \tau\}.
\end{equation}

\noindent
We repeat this calculation across thresholds from $\tau=0$ to $\tau=100$. To account for uncertainty in the risk index, we add Gaussian noise with standard deviation $\sigma=5.0$ to each road's $\mu$ value and run 100 Monte Carlo iterations at every threshold.

At each threshold, we track four outcomes. The first is the size of the giant connected component, measured as the number of nodes that remain in the largest connected part of the network. The second is the population in true isolates: city nodes with degree zero after road removal. The third is the number of connected components. The fourth is the share of predicted flow between cities that uses removed roads. Together, these measures separate local isolation from wider fragmentation: one asks who is cut off immediately, while the other asks when the road system stops functioning as a connected continental network.

\subsection*{Ethics and data protection}
This study analyzes public, previously published, or aggregate third-party data. It does not involve intervention with human participants, recruitment, surveys, experiments, or access to individual-level personal records. Migration flows are analyzed only at aggregate directed country-pair/year scale from the cited source, and no individual-level Facebook data are accessed, stored, or redistributed. In our assessment, the study poses no risks to individuals or individual privacy.

\begin{figure}[!ht]
\centering
\includegraphics[width=\textwidth]{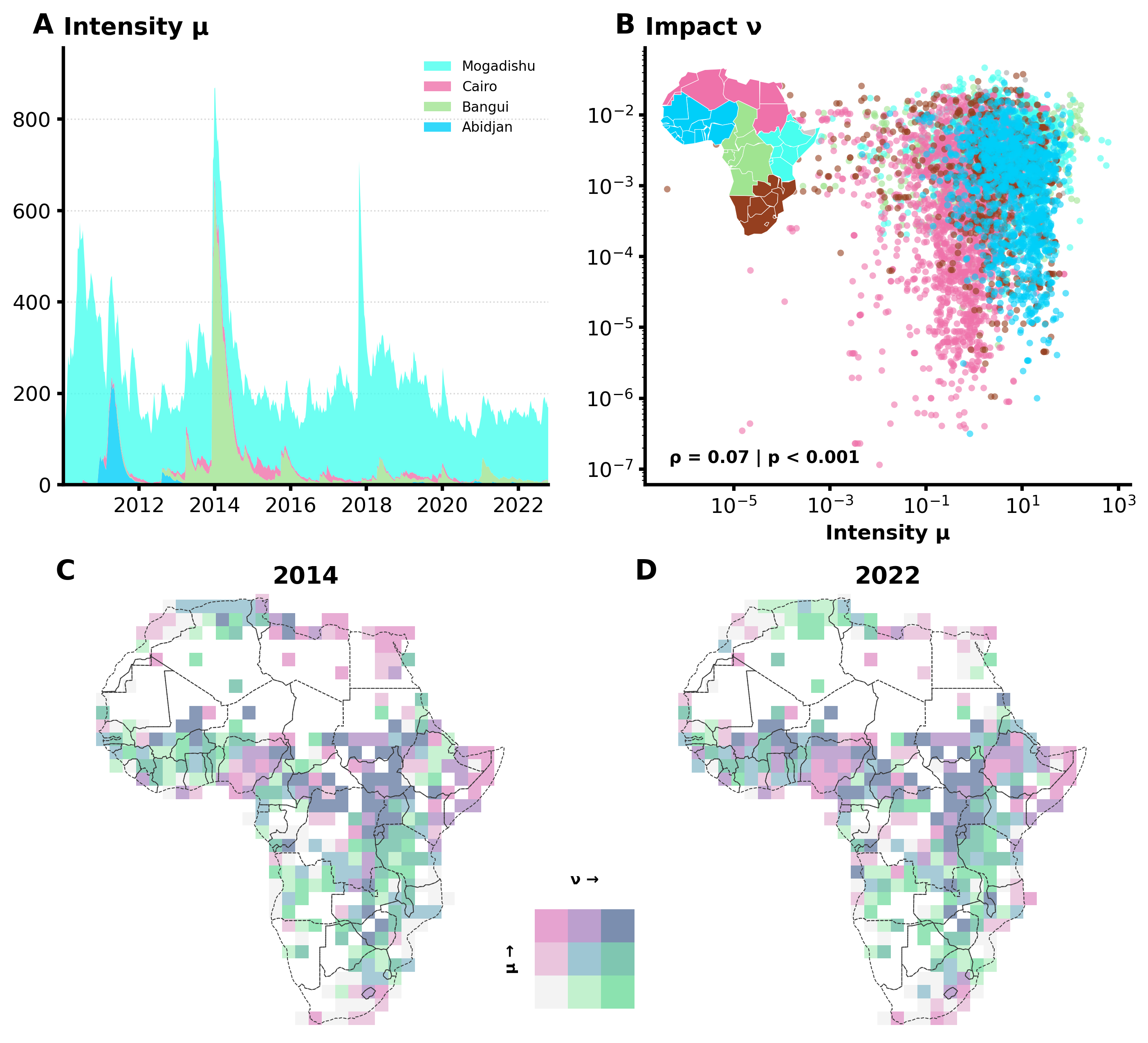}
\caption{\textbf{Conflict and projected movement overlap weakly but consequentially.} A: City conflict intensity is persistent in Mogadishu and more episodic in Cairo, Bangui, and Abidjan, showing why a surface of recurring risk is more informative than a single event count. B: Road conflict intensity $\mu$ and projected movement from radiation $\nu$ are only weakly related ($\rho = 0.066$, $p<0.001$), so roads that carry large flows are not automatically roads with high risk. C and D: The bivariate maps locate the exceptions: some cells combine high conflict and high projected movement, but much of the network sits off that diagonal.}
\label{fig1}
\end{figure}

\section*{Results}
The African road network in our analysis connects 2,162 cities through 9,159 major roads, 5,199 intermediate road junctions, and 274 border crossings; construction details are in Supplementary Section~\nameref{si:network_all_highways}. It links an urban population of 460.7 million people, but not evenly: West Africa contains 584 cities and 134.3 million urban residents, while Southern Africa contains 177 cities and 33.6 million. The network therefore begins with a basic asymmetry. Some regions have many cities, many people, and many roads; others have fewer alternatives if one route becomes unsafe.

Conflict is more concentrated than the road network itself. We estimate conflict intensity $\mu$ with a Hawkes-style memory kernel, using the decay calibration in Supplementary Fig.~\ref{S2_Fig}. On the edge-risk surface used for road-level analyses, $\mu_e$ ranges from near zero to 613.6, with a mean of 7.97 and a median of 1.88. City-level peaks are larger because they summarize the node surface rather than the road-segment index: the largest city peaks appear in Juba, Mogadishu, Maiduguri, and Bangui, placing intense violence in East, West, and Central Africa rather than in a single regional pocket. Cities with the highest average intensities are summarized in Table~\ref{tab:top_city_mu}, separating persistent exposure from acute peaks. Between 2015 and 2022, the number of cities with nonzero conflict intensity grew from 1,000 to 1,490, a 49\% expansion of the urban conflict footprint; Supplementary Fig.~\ref{S3_Fig} shows the wider temporal spread.

\begin{figure}[!ht]
\centering
\includegraphics[width=\textwidth]{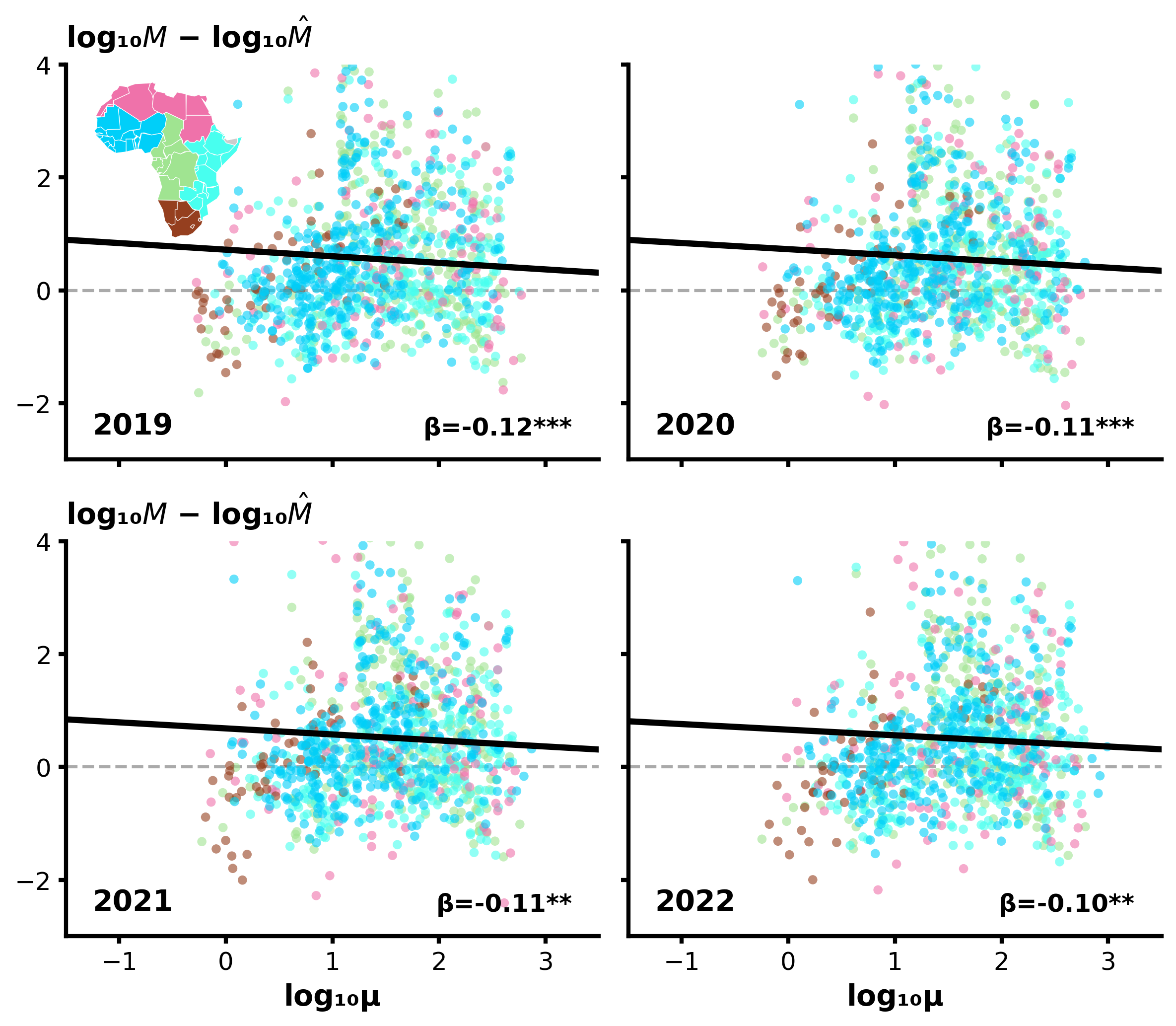}
\caption{\textbf{Migration data support the corridor exposure measure.} Each panel compares observed migration between country pairs, measured as $\log_{10}(M/\hat{M})$ relative to the radiation prediction, with conflict intensity along the weighted road corridors connecting many cities in the two countries. The annual slopes are negative in every year from 2019 to 2022, with coefficients from -0.12 to -0.10 and directed country-pair clustered standard errors. In the current main specification, a doubling of corridor conflict is associated with about 7\% lower bilateral migration, and a tenfold increase with about 22\% lower migration.}
\label{fig2}
\end{figure}

\subsection*{Conflict reaches corridors that organize movement}
Next we assess whether roads that experience conflict are peripheral or critical--whether they sit on routes that organize movement between cities and countries. Fig.~\ref{fig1}\textbf{B} compares conflict intensity $\mu$ with projected movement from radiation $\nu$ on each road. The two measures are only weakly related. Many violent corridors carry little expected movement, and many corridors that carry large flows have low recent conflict; the vulnerable corridors are the smaller set where high conflict and high projected flow meet. Those corridors are the places where violence is most likely to matter beyond the road where it occurs.

The regional estimates sharpen this distinction, which we show in Table~\ref{tab:mu_nu_regional}. The weak continental relationship is not a uniform pattern: Central Africa has the clearest positive coupling, North Africa is weakly positive, East Africa is not distinguishable from zero, and Southern and West Africa are negative in this simple comparison at the level of roads. The same value of $\mu$ therefore means different things depending on where it sits in the road system.

The maps in Fig.~\ref{fig1}\textbf{C} and \textbf{D} make the same asymmetry spatial. In both periods, 2014 and 2022, cells with large flows and cells with high conflict do not collapse into one geography. Some places combine the two, especially around exposed corridor systems, but many areas with large flows remain comparatively less exposed and many areas with high conflict sit on parts of the network with lower projected flow. Conflict risk is therefore a matter of both position and intensity.

\subsection*{Migration data ground the corridor interpretation}
We use migration flows here for a specific purpose. The paper is about road network vulnerability, not migration as an outcome in its own right. The migration analysis asks whether spatial interaction is partly organized by the road network, and whether conflict along the resulting corridors moderates that interaction. If observed movement did not line up with the radiation model routed through roads, the percolation analysis would rest on a weaker mobility model. Fig.~\ref{fig2} therefore tests the corridor exposure measure against observed bilateral movement before we turn to network disruption.

Migration data combine at least two responses to conflict: violence can push people away from the site of an event, increasing movement from affected places, but it can also make a corridor harder to use, decreasing movement through the route it touches. We use corridor exposure as a suggestive test of this second channel, asking whether conflict along the road path between two countries predicts lower migration after accounting for the movement expected from population, distance, and intervening opportunities. Across 4,071 observations of country pairs from 2019 to 2022, higher conflict intensity along weighted corridors between cities is associated with lower bilateral migration after controlling for expected flows under radiation. The annual estimates are consistently negative, implying that a doubling of corridor conflict is associated with about 7\% lower migration and a tenfold increase with about 22\% lower migration in the current main specification; we report the corresponding regional estimates in Table~\ref{tab:conflict_migration_regional_breakdown}.

The regional estimates qualify the pooled result, which we show in Table~\ref{tab:conflict_migration_regional_breakdown}. The negative association is clearest for routes crossing between regions and for routes involving North and West African origins, while routes within the same region are not consistently negative. Southern Africa is an exception in this simple regional split, where the coefficient is positive and the sample is small; we therefore treat the regional estimates as heterogeneity checks rather than as separate regional mechanisms. The result for routes crossing regions points toward a corridor interpretation, but it should not be read as fully separating route exposure from origin and destination conditions. In stricter specifications based on prior migration choices, lagged migration absorbs much of the variation in corridor choice, leaving the evidence on route exposure suggestive rather than definitive.

This migration exercise grounds the findings that follow more than it claims migration as the paper's endpoint. It shows that spatial interaction broadly follows the corridor structure of the road network, and that conflict along those corridors is associated with lower movement. With that empirical footing, we turn to the network question: what happens when roads exposed to conflict fail?

\begin{figure}[!ht]
\centering
\includegraphics[width=\textwidth]{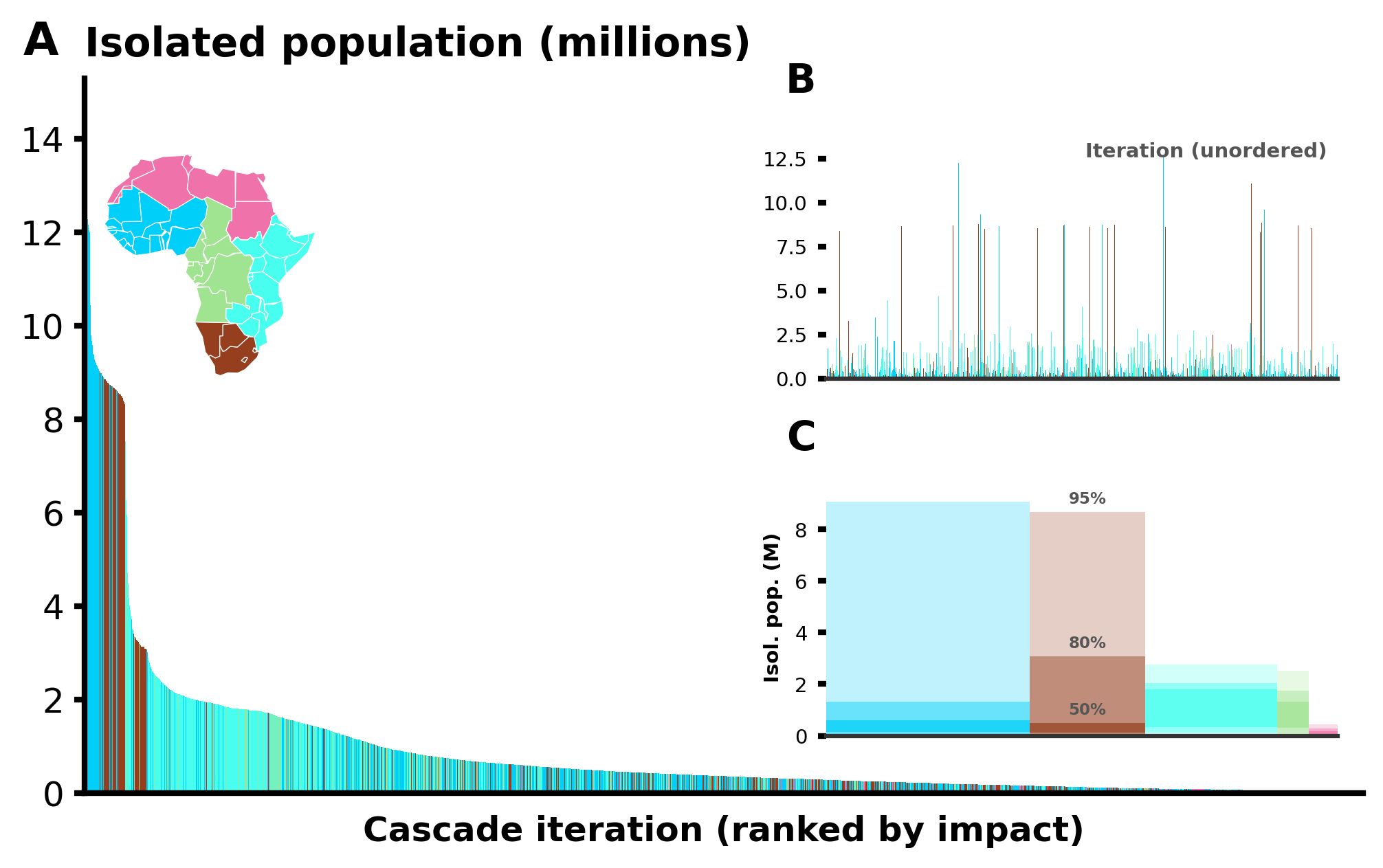}
\caption{\textbf{Conflict cascades are usually small, but the upper tail is large.} We seed failures on roads exposed to conflict and allow every other road to fail with probability that decays with distance and rises with its own conflict intensity. A: Ranked nonzero simulations show a long tail: most cascades isolate modest populations, but 18.6\% of all simulations isolate more than one million people and the largest isolates 13.3 million. B: The same simulations in their original order show that large events are intermittent rather than gradual. C: Regional summaries combine frequency and severity: each block's width counts nonzero cascades, while nested bands show the central 50\%, 80\%, and 95\% of outcomes for isolated population. West Africa produces many large cascades, East Africa has a higher typical impact among nonzero events, Southern Africa has a wide upper tail, and North Africa remains comparatively insulated.}
\label{fig3}
\end{figure}

\subsection*{Local cascades are usually small, but a heavy tail threatens integration}
\begin{figure}[!ht]
\centering
\includegraphics[width=\textwidth]{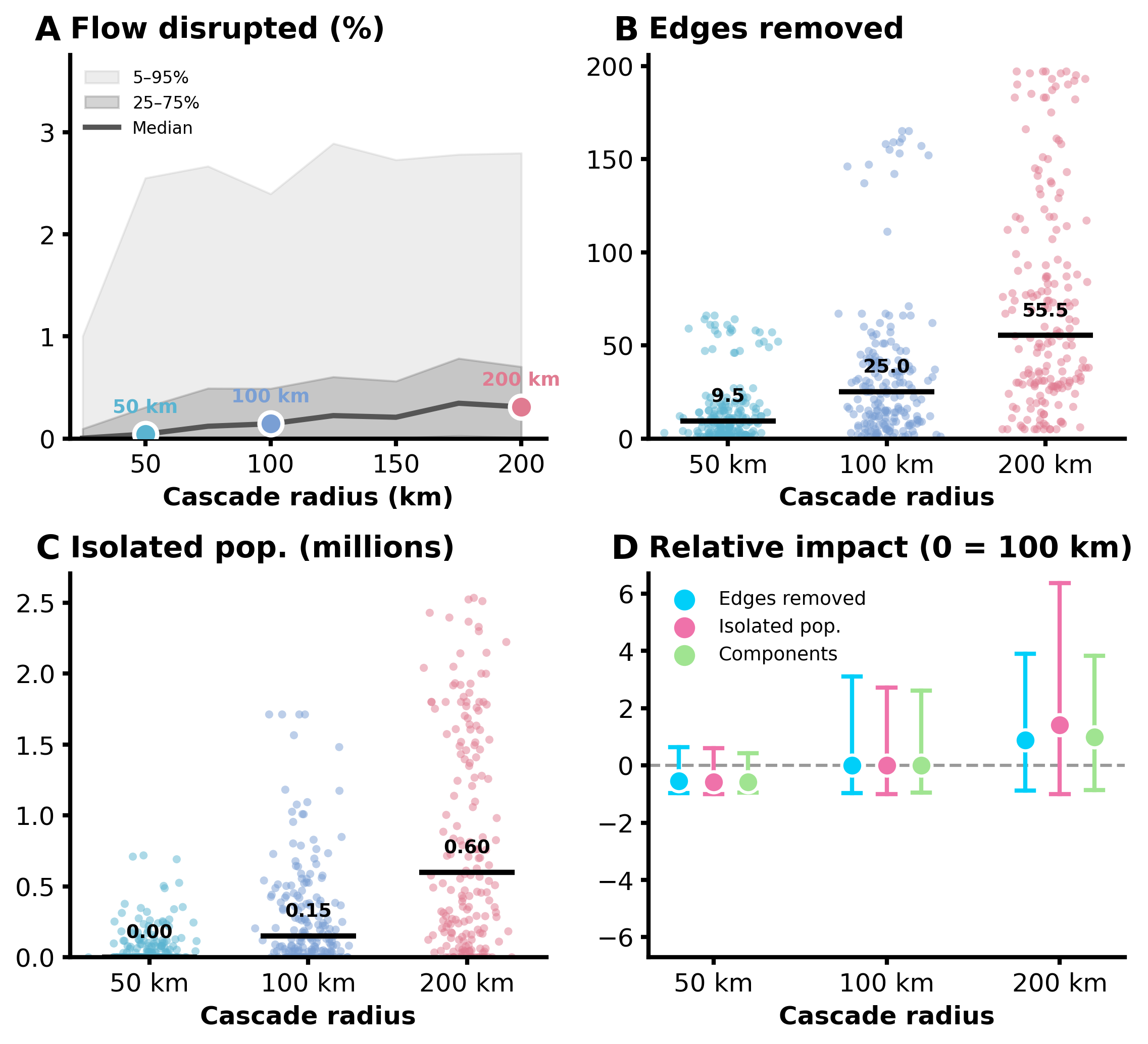}
\caption{\textbf{Cascade range expands typical damage more than tail risk.} We vary the hard radius used in the sensitivity check. A: The median share of modelled flow disrupted rises from 0.04\% at 50 km to 0.14\% at 100 km and 0.31\% at 200 km, while the 5--95\% band opens quickly and then changes little. B and C: Wider radii remove more roads and isolate larger populations, with median removed edges increasing from 9.5 to 25.0 to 55.5 across 50, 100, and 200 km and median isolated population rising from near zero to 0.15M to 0.60M. D: Relative impacts are centered at the 100 km specification; the upper whiskers overlap across outcomes, suggesting that the most damaging cascades depend more on hitting corridors with few substitutes than on additional spatial reach.}
\label{fig4}
\end{figure}

The migration analysis grounds the modelled corridor flows at the level of routes; clustered road failures ask how far a local disruption can travel before it becomes a regional problem. We model localized conflict cascades using a contagion kernel with a power law, which we show in Fig.~\ref{fig3}. A seed edge is selected with probability proportional to its conflict intensity $\mu_e$, and every other edge can fail with probability $1 - \exp(-\mu_f / d^{\alpha})$, where $d$ is the Euclidean distance between edge midpoints in kilometres, floored at 1\,km, $\mu_f$ is the receiving edge's conflict intensity, and $\alpha = 1.1$. The main cascade is not radius-capped; Fig.~\ref{fig4} reports the hard-radius sensitivity.

Most cascades are limited. The median cascade isolates 244,000 people, modest relative to the continent's 461 million urban residents. But the upper tail changes the interpretation: nearly one in five cascades isolates more than 1 million people, and the largest isolates 13.3 million people in West Africa through the removal of 68 road segments. Seed risk is only weakly correlated with cascade impact, confirming that outcomes depend as much on where the disruption occurs as on how severe the triggering event is. We report the full tail probabilities in Table~\ref{tab:cascade_region_tail}.

The regional breakdown clarifies the tail, which we show in Table~\ref{tab:cascade_region_tail}. West Africa produces the largest number of nonzero cascades, but East Africa has the higher average impact per event, consistent with a sparser road network with fewer alternative routes. Southern Africa contributes less often, but its bottleneck roads create a wide upper tail: a cascade seeded in the wrong place can sever connections for more than 10 million people. North Africa is comparatively insulated, reflecting both lower recent conflict intensity and a denser northern road system.

The sensitivity of these results to cascade radius appears in Fig.~\ref{fig4}. Wider cascades remove more roads and disrupt more total flow on average, as expected. The upper tail changes less. Once the cascade radius reaches roughly 100\,km, the 95th percentile of disrupted flow stabilizes, suggesting that the most damaging outcomes are driven by topology more than by spatial reach. Once a cascade reaches a major corridor, additional distance adds little marginal disruption.

\begin{figure}[!ht]
\centering
\includegraphics[width=\textwidth]{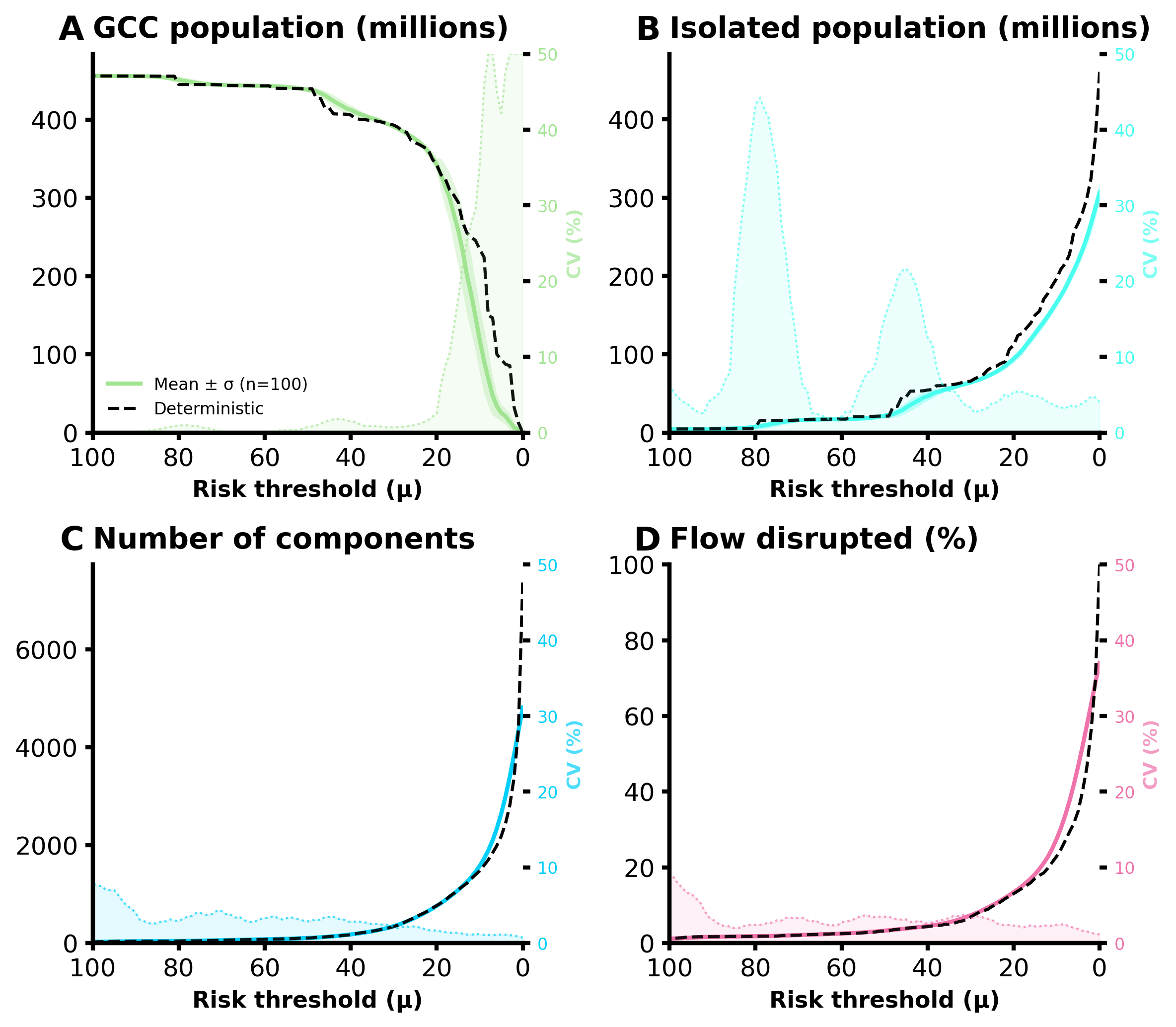}
\caption{\textbf{Attack percolation reveals a sharp fragmentation threshold.} We remove roads whose edge conflict intensity exceeds a threshold and compare deterministic removal (black dashed) with 100 noisy runs (colored mean and ribbon); the dotted line on the right axis gives the coefficient of variation. A: The giant connected component remains nearly intact when only roads with the highest risk are removed, but falls through the critical zone between $\mu_e=20$ and $\mu_e=10$. B: Isolated population rises across the same zone. C and D: Components and disrupted flow accelerate as removal moves from rare roads with high conflict into the broader exposed network. In that transition, 874 additional roads, 9.5\% of the network, move the GCC from 85\% to 39\%.}
\label{fig5}
\end{figure}

\subsection*{Accumulated corridor risk fragments the network sharply}
Our approach to measuring local cascades asks how much damage local disruptions can do when they are contained but clustered; percolation asks when accumulated corridor risk begins to fragment the whole network. In Fig.~\ref{fig5}, we remove all edges whose conflict intensity $\mu_e$ exceeds a threshold and measure the resulting fragmentation of the road network. To account for measurement uncertainty in edge risk, we add Gaussian noise ($\sigma = 5.0$) across 100 Monte Carlo iterations per threshold. The network holds together when only the most extreme roads are removed: even after removing roads with $\mu_e > 30$, the giant connected component (GCC) retains 93.7\% of nodes.

The transition, however, is sharp. The critical zone lies between $\mu_e = 20$ and $\mu_e = 10$: the removal of just 874 additional roads---9.5\% of the total network---tips the GCC from 85\% to 39\%. Near that threshold, small differences in which roads are classified as high risk produce very different fragmentation outcomes. This is the structural version of the corridor problem. Most roads are low risk, but the network's fate is determined by the remaining third, where conflict exposure and sparse alternatives begin to overlap. The full threshold series is reported in Table~\ref{tab:percolation_thresholds}.

\begin{figure}[!ht]
\centering
\includegraphics[width=\textwidth]{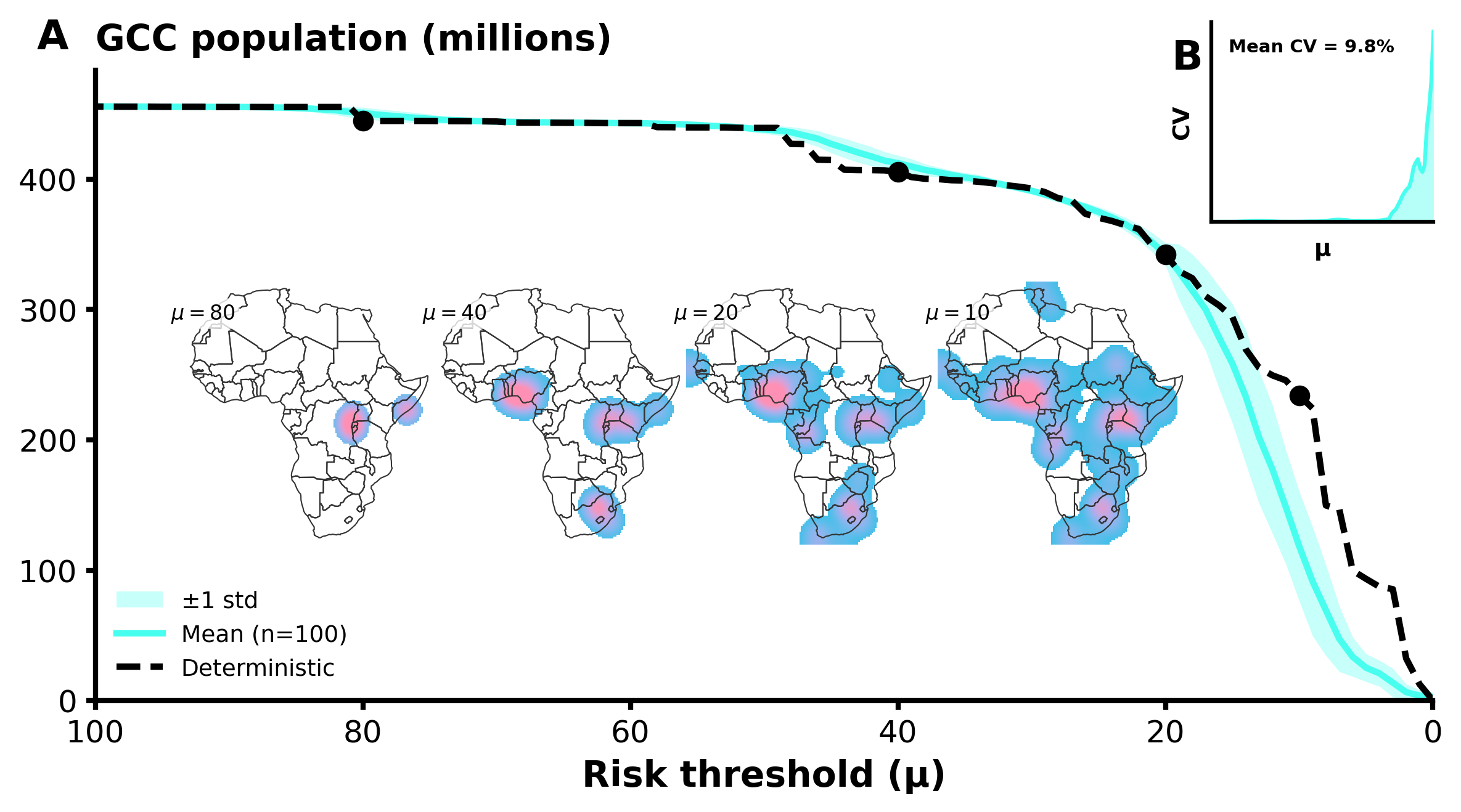}
\caption{\textbf{Fragmentation begins in sparse corridors with high risk before spreading across regions.} A: The threshold curve tracks population remaining in the giant connected component, while inset maps show concentrations of isolated cities weighted by population at selected thresholds. Isolation is negligible at high thresholds, appears first in East Africa around Somalia and South Sudan, and then spreads toward West Africa and the Sahel as the threshold falls. B: The coefficient of variation stays low while only extreme roads are removed, then spikes near the percolation zone, where small changes in edge risk determine which parts of the continent disconnect.}
\label{fig6}
\end{figure}

The geography of this fragmentation appears in Fig.~\ref{fig6}. At high thresholds, isolation is negligible. As the threshold falls, the first clusters appear in East Africa---particularly Somalia and South Sudan, where high conflict intensity coincides with a sparse road network. Fragmentation then extends toward the Sahel corridor in Mali, Burkina Faso, and Niger, a second zone of vulnerability in a region with 134.3 million urban residents. The geographic progression---from the Horn of Africa to the Sahel to wider continental fragmentation---suggests that the network's structural weaknesses are aligned with its most active and expanding conflict zones.

Supplementary Fig.~\ref{S8_Fig} expands Fig.~\ref{fig6} across 2019--2022, showing that the exposed clusters move with the yearly conflict surface but recur along the same sparse corridors. Supplementary Fig.~\ref{S9_Fig} shows a second margin of risk: even when the loss of roads exposed to conflict does not fragment the network, rerouting can still impose long journeys. Among roads with finite substitutes, removal adds a median 142 km to the route---roughly 2.4 hours at 60 km/h, before checkpoints, borders, or poor road conditions are counted. Among roads that also carry projected movement and conflict, the median added distance is 181 km. Another 2,339 roads have no substitute at all, including 1,092 in North Africa. Supplementary Table~\ref{tab:top_vulnerable_corridors} lists examples of roads where conflict intensity, projected movement, and low redundancy overlap. The risk comes from rare tail events that disrupt major corridors and from prolonged journeys where conflict forces movement onto weaker substitutes.

\section*{Discussion}
Road transport carries most movement in Africa, including roughly 80\% of goods traffic and 90\% of passenger traffic. Yet many corridors operate with thin redundancy: in 2005, only 22.7\% of the African road network was paved, and many routes remain exposed to border delays, high transit tariffs, market access barriers, and transport restrictions \cite{UNTransportAfrica09, AfricaEUCoop-Transport}. Conflict enters this system as more than an external shock. Armed groups can exploit sparse infrastructure, seize towns, and disrupt roads in ways that deepen existing transport constraints \cite{prieto2020uncovering}.

Our central finding is that risk and importance do not sit in the same places. Some roads matter because they carry movement; others matter because they accumulate violence; the vulnerable corridors are the ones where those two geographies meet. North Africa contains many corridors that carry large flows with lower recent conflict, Somalia and other isolated parts of the network contain conflict on roads with lower projected flow, and the Sahel forms a broader band where risk and projected movement begin to overlap.

Isolation is the human consequence of this geography. In the cascade simulations, large failures are more likely in West and East Africa; in the percolation analysis, removing exposed roads disrupts a larger share of projected flow in West Africa, where detours or cancelled journeys would mean longer travel times, added delays, and higher emissions. Planning for resilience should therefore ask whether the network has enough substitutes to keep people and goods connected while damaged roads are repaired \cite{bruneau2003framework, roy2019quantifying}.

We observe bilateral migration and modelled route exposure, not the individual journeys that were cancelled, delayed, or rerouted. ACLED also depends on reported events, and the road network abstracts away local road quality, checkpoints, seasonal conditions, and informal transport options. These limits make the migration evidence a grounding check on corridor exposure rather than a complete account of mobility under conflict. The network result remains the central point: consequences are largest where exposed roads have few substitutes.

Seeing conflict through the network changes the policy question. Areal measures can tell us where violence occurs, but they cannot show which journeys depend on the affected routes; the network view lets us ask how a local event changes access, detours, and isolation across the wider system \cite{harari2018conflict}. That matters for infrastructure strategy. The result is not simply to protect the busiest roads. It is to protect high-risk, low-redundancy corridors where movement and exposure overlap.

\clearpage

\section*{Supporting information}

\subsection*{The network of all African highways}
\phantomsection
\label{si:network_all_highways}

{

We construct the African highway network from OpenStreetMap primary roads, highways, and trunk roads \cite{OpenStreetMap}. Africapolis provides the location and population of urban agglomerations with more than 100,000 inhabitants, along with smaller cities near major roads that are needed to keep the network connected \cite{Africapolis}. Road networks require an additional construction step because their nodes are not as naturally defined as train stations in a railway network. We therefore represent both cities and road junctions as nodes.

We spatially align cities and roads, integrate them into a single network, and join disconnected components based on road proximity \cite{prieto2022constructing, prieto2022detecting}. We then simplify the network by dissolving degree-2 nodes that connect the same road type at both ends, replacing the two edges with one longer edge whose length is their sum. The resulting network contains 9,159 edges and 7,361 nodes: 2,162 cities and 5,199 road junctions. Edges represent highway segments between nodes. Road type is used to estimate speed and travel time. Edges whose endpoints lie in different countries are treated as border crossings and receive an additional 120 minutes of travel time; travel through an urban agglomeration receives an additional time cost proportional to the city's population. Each edge therefore has a length, border status, whether it intersects a city, and estimated travel time.

}

\clearpage

\subsection*{ACLED coverage and reporting bias}
\phantomsection
\label{si:acled_bias}

{
ACLED records violence, conflict, civil unrest, and other forms of social upheaval, but it is a record of reported events. Coverage can vary with media density, source networks, population, and the political visibility of violence. We therefore check for reporting bias rather than treating ACLED as a perfect census, asking whether it recovers a coherent conflict signal at the scale of countries, cities, and roads.

\setcounter{figure}{0}
\renewcommand{\thefigure}{S\arabic{figure}}

\refstepcounter{figure}
\paragraph*{\thefigure\ Fig.}
\label{S1_Fig}
\begin{center}
\includegraphics[width=0.95\textwidth]{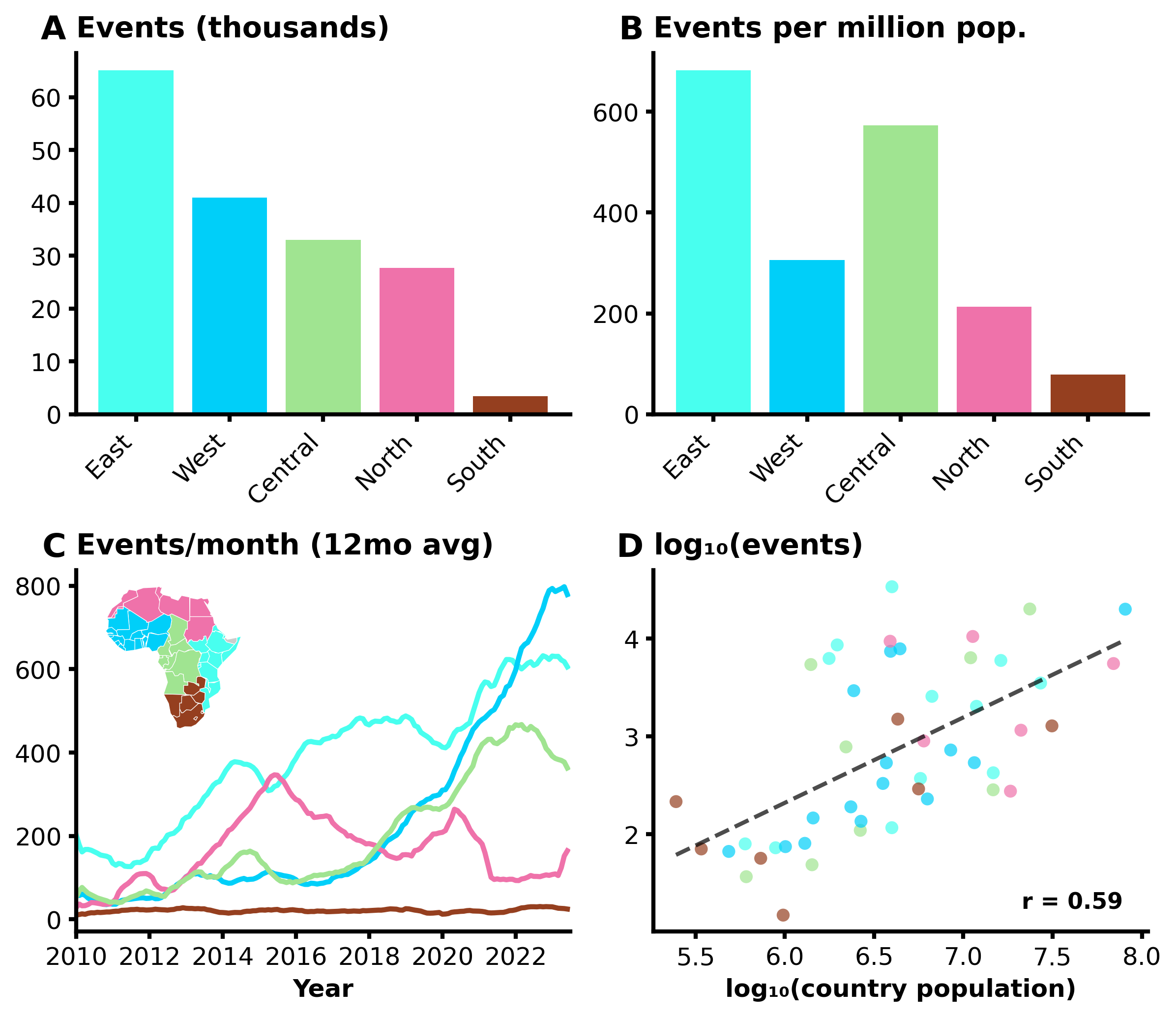}
\end{center}
\textbf{ACLED coverage is uneven, but not arbitrary.} A and B: Event counts vary strongly by region, both in absolute terms and per million residents; East Africa has the most reported events and the highest rate adjusted for population, while Southern Africa is low on both scales. C: Rolling averages over 12 months show real temporal structure rather than flat reporting, with East and West Africa rising sharply after 2020 and Central Africa increasing more gradually. D: Country event counts rise with population ($r=0.59$), which is consistent with both exposure and reporting capacity. These patterns are the expected limitations of reported conflict data. They are also why we use ACLED as a consistent record of reported political violence, not as a claim that every unreported event is observed.

\clearpage

\subsection*{Hawkes-style conflict memory and its parameters}
\phantomsection
\label{si:hawkes_parameters}

We use a Hawkes-style self-exciting intensity index to model conflict memory in city $i$. The index $\mu_i(t)$ at time $t$ is
\begin{equation}
\mu_i(t) = \sum_{s:t \geq t_s} \kappa_s \exp{ -\phi (t - t_s)},
\end{equation}
where $t_s$ is the date of event $s$, $\kappa_s$ is the event weight, and $\phi > 0$ controls memory decay. Smaller values of $\phi$ allow an event to remain influential for many days; larger values make the intensity respond mainly to recent events. The sum includes only events that occurred before time $t$, and each contribution fades as the time since the event grows. We use this kernel as an exposure measure, not as a fully estimated generative Hawkes process with fitted background and triggering terms.

We take $\kappa_s = 1 + f_s$, where $f_s$ is the number of event fatalities, so that nonfatal events still contribute to intensity. We estimate $\phi$ by asking which decay rate gives the most information about violence over the next two weeks. For each test value of $\phi$, we compute the Hawkes intensity for city $i$ and correlate $\mu_i(t)$ with $E_i(t)$, the average daily impact experienced in the same city from $t$ to $t+14$. A higher correlation means that current intensity better predicts the next two weeks of violence. Varying $\phi$ gives the calibration curve $\theta(\phi)$; the audited calibration gives $\phi^\star = 0.055$, corresponding to a half-life of approximately 13 days (Supplementary Fig.~\ref{S2_Fig}).

\refstepcounter{figure}
\paragraph*{\thefigure\ Fig.}
\label{S2_Fig}
\begin{center}
\includegraphics[width=\textwidth]{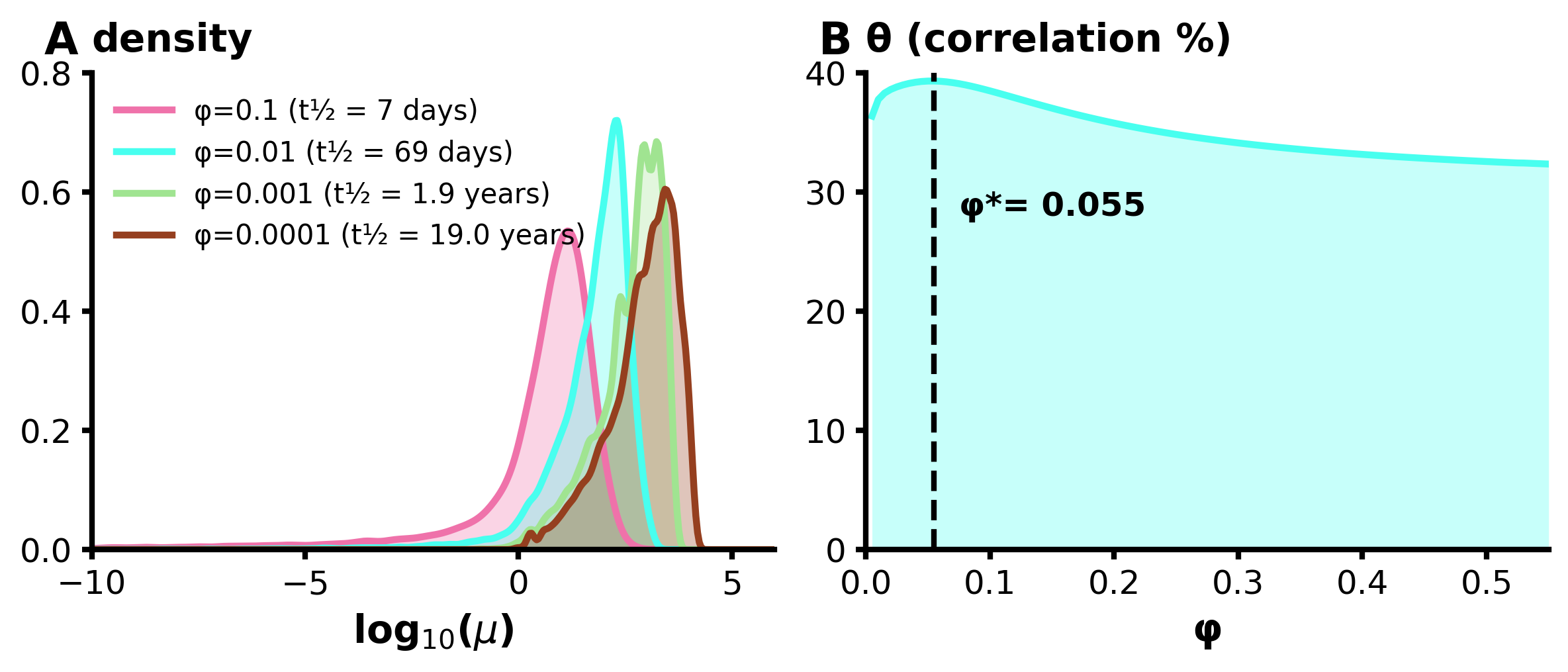}
\end{center}
\textbf{Conflict memory has two useful time scales.} A: Varying $\phi$ changes how strongly recent events dominate the Hawkes intensity distribution: fast decay isolates the most recent shocks, while slower decay keeps older violence in the corridor's risk history. B: The predictive calibration asks which decay rate best anticipates fatalities in the next 14 days. The audited curve peaks at $\phi^\star = 0.055$ ($\theta = 0.393$), a half-life of approximately 13 days. We use that short memory calibration for conflict forecasting, cascades, and percolation, while the migration robustness checks separately test longer corridor memories.

\clearpage

\refstepcounter{figure}
\paragraph*{\thefigure\ Fig.}
\label{S3_Fig}
\begin{center}
\includegraphics[width=0.95\textwidth]{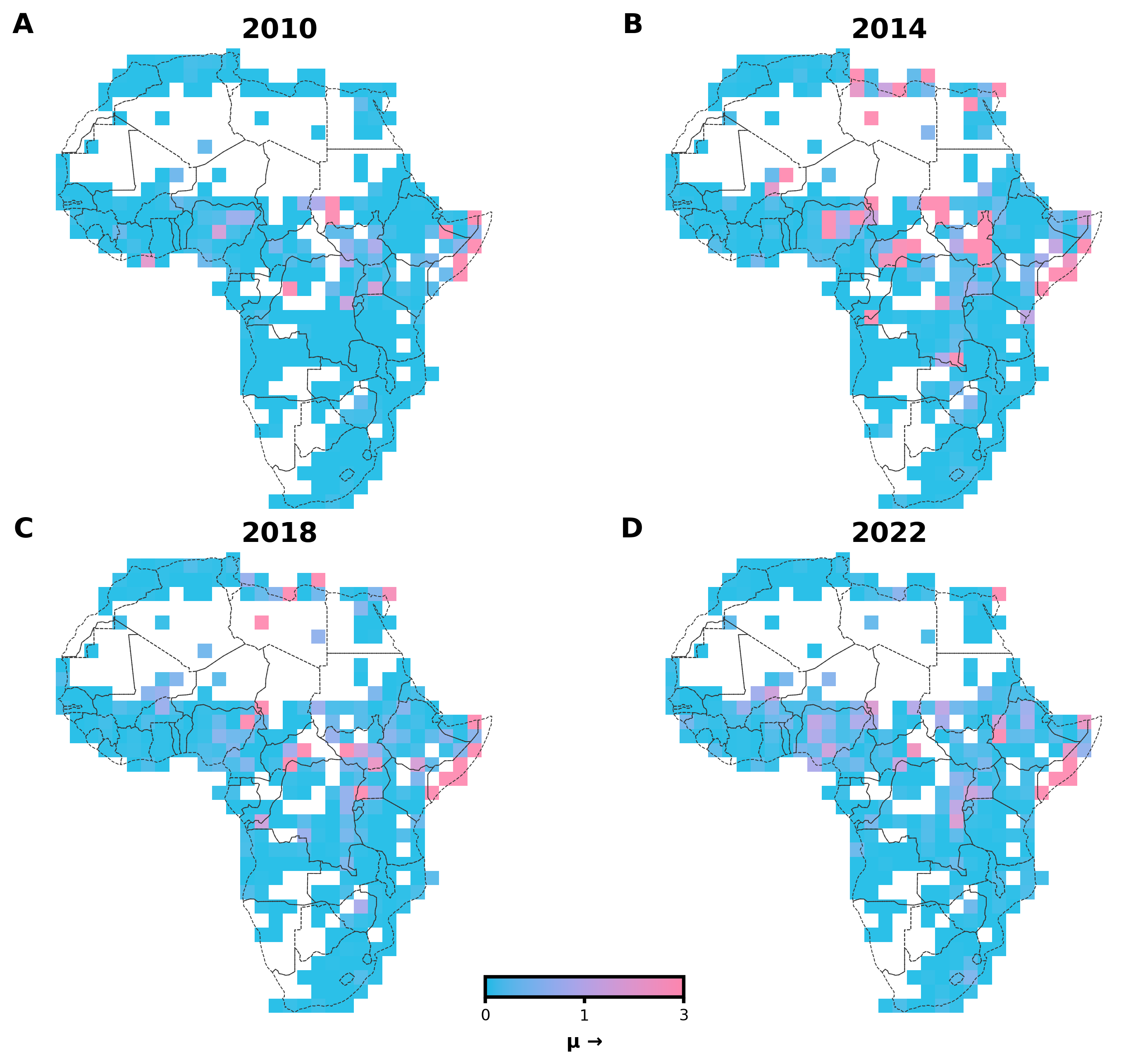}
\end{center}
\textbf{The urban conflict surface expands across more corridors.} Grid cells weighted by population summarize Hawkes conflict intensity in 2010, 2014, 2018, and 2022. The early map is sparse; by 2018 and 2022, recurrent intensity covers a wider set of urban corridors, especially in the Sahel, the Horn of Africa, and parts of Central Africa. The figure supports the main text claim that the network is exposed to a growing conflict footprint, beyond a few isolated events with many fatalities.

\clearpage

\subsection*{Gravity and radiation models}
\phantomsection
\label{si:gravity_radiation_models}

Modelled movement enters the paper in two linked ways. First, flows between cities are routed through the road network to create $\nu_e$, the projected flow measure that tells us which roads organize expected travel between cities. Second, the same latent flows are aggregated to country pairs to build the expected flows in a regression predicting migration with conflict. This gives us a way to ground a central assumption of the paper--that conflict and spatial interaction are connected---in empirical data. Observed bilateral migration is therefore both a control variable for Fig.~\ref{fig2} and the empirical anchor for the projected flow measure used throughout the network analysis.

This anchor is imperfect, because international migration is only one form of movement, but research indicates that migration and trade are coupled \cite{white2011international}. It gives us a measured pattern against which to check whether the model places movement in plausible corridors. Radiation and calibrated gravity agree on the broad road backbone: among roads with nonzero projected flow under either model, their $\nu$ values at the level of roads have a Spearman correlation of $\rho = 0.74$. They differ enough, however, that the choice of main model matters. Radiation has the higher annual correlation with observed migration shares in every year, ranging from 0.51 to 0.54 compared with 0.47 to 0.50 for gravity. Gravity has lower log RMSE in every year, ranging from 1.12 to 1.19 compared with 1.30 to 1.32 for radiation. We therefore use radiation as the main baseline because it better ranks migration corridors, and use calibrated gravity as a robustness model because it better matches absolute scale.

\refstepcounter{figure}
\paragraph*{\thefigure\ Fig.}
\label{S4_Fig}
\begin{center}
\includegraphics[width=0.95\textwidth]{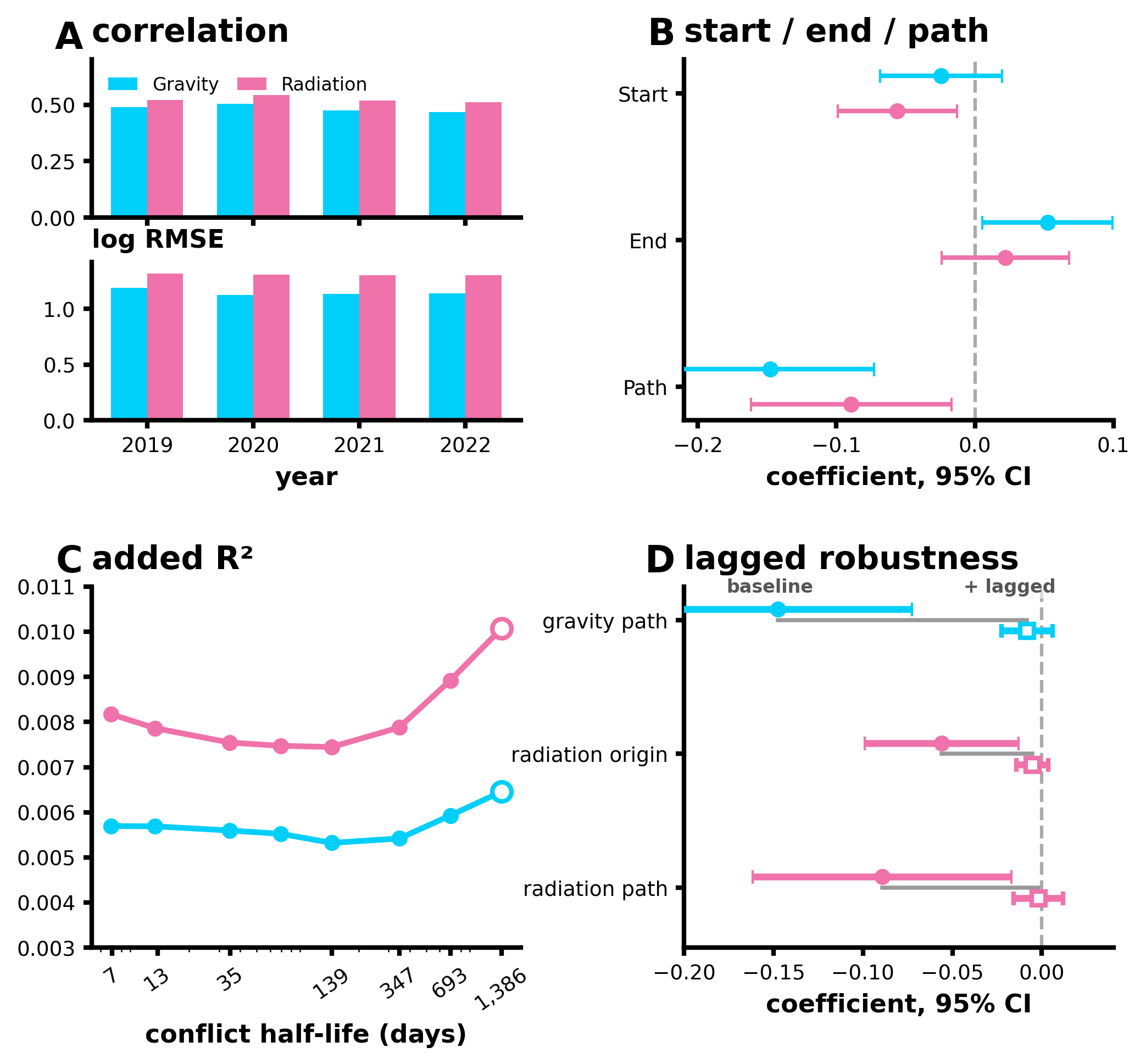}
\end{center}
\textbf{Radiation better ranks observed migration corridors, while gravity better matches absolute scale.} A: Radiation has the higher annual correlation with observed migration shares, while gravity has lower log RMSE. We use radiation for the main projected flow measure and migration regression because the comparison depends on whether corridors are high or low relative to expected movement, and retain gravity as the robustness baseline with lower error. B: When origin, destination, and interior path conflict are separated, interior path conflict is negative under both model families. C: Longer conflict memory improves the migration fit, especially for the radiation baseline. D: Adding migration from the prior year attenuates the conflict coefficients, showing that revealed corridor preference absorbs much of the persistent route signal.

\medskip
Because gravity remains the main robustness baseline, we calibrate its distance decay rather than choose a value by convention. For each year, we route gravity flows between cities through the road network, aggregate them to country pairs, and select the $\beta$ that maximizes the correlation between logged predicted and observed migration shares. The accepted range is narrow: $\beta^\star = 1.3$ in 2019 and $\beta^\star = 1.2$ from 2020 to 2022, with accepted values between 1.0 and 1.5 across the four years.

The calibrated gravity model is informative but incomplete. Its annual correlations with observed migration range from 0.47 to 0.50, and the fitted slopes are below one, meaning the model compresses the most extreme migration corridors. That is enough to make gravity an informative sensitivity check. It is not enough to make it the main baseline when radiation better preserves the rank ordering of observed migration.

\refstepcounter{figure}
\paragraph*{\thefigure\ Fig.}
\label{S5_Fig}
\begin{center}
\includegraphics[width=0.95\textwidth]{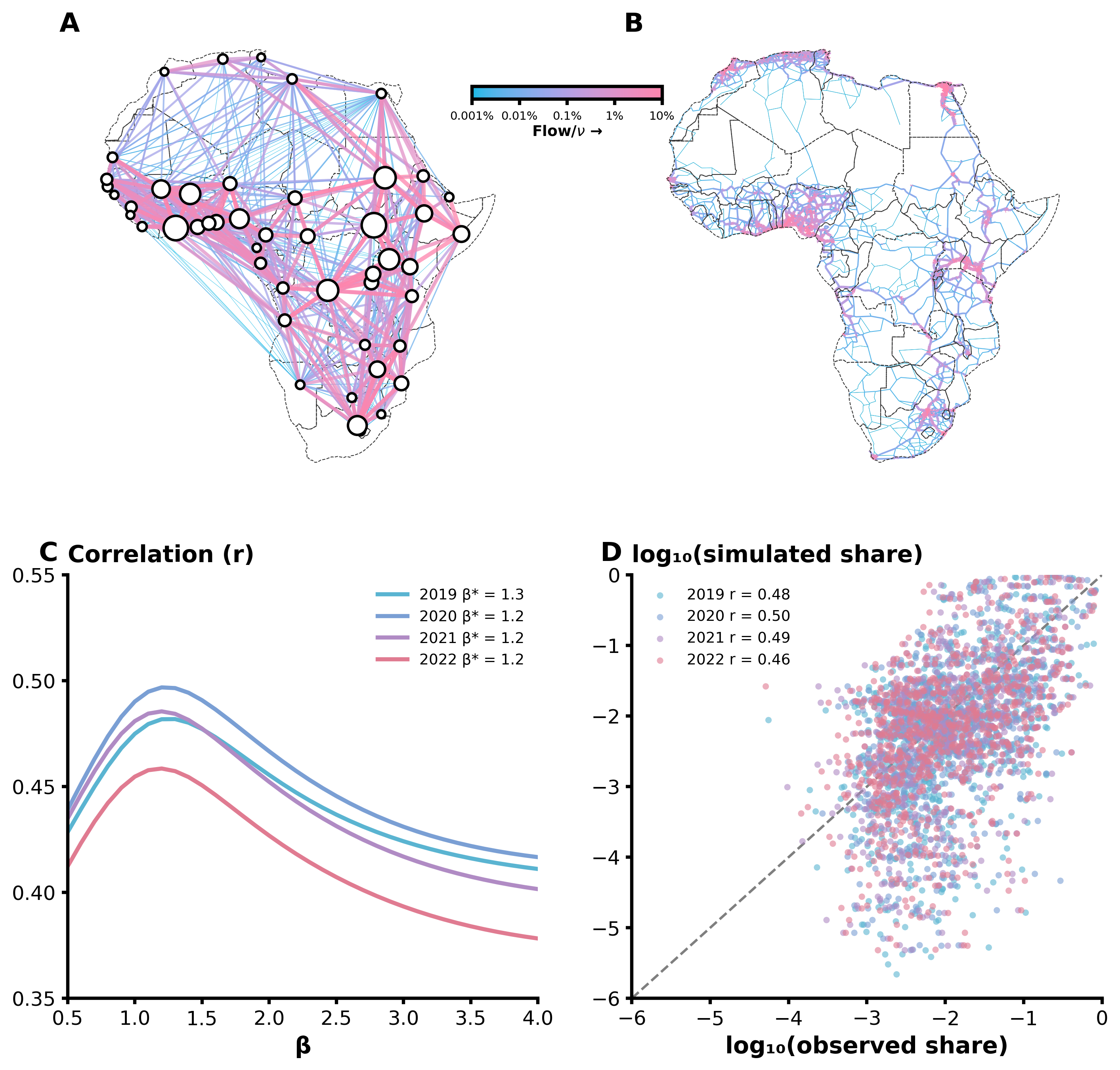}
\end{center}
\textbf{Gravity calibration provides a robustness baseline with calibrated distance decay.} A and B connect the migration data between countries to the gravity flows between cities on the road network used for the robustness specification. C: The calibration is stable across years, selecting $\beta^\star = 1.3$ in 2019 and $\beta^\star = 1.2$ from 2020 to 2022. D: Aggregated gravity predictions recover a meaningful but incomplete share of observed migration. The calibration supports gravity as a sensitivity check for error in absolute flows, while the model comparison in Supplementary Fig.~\ref{S4_Fig} supports radiation as the main baseline for rank ordering \cite{chi2025measuring}.

\medskip
Swapping gravity for radiation changes the interpretation more than the sign of the result. In the pooled decomposition that separates endpoints from paths at the main migration memory ($\phi = 0.0005$), interior path conflict is negative under both baselines. The gravity version gives a larger coefficient for interior path conflict ($-0.148$, $p<0.001$), while origin conflict is not distinguishable from zero. The radiation version gives a smaller but still negative coefficient for interior path conflict ($-0.089$, $p=0.016$) and also retains a negative term for origin conflict ($-0.056$, $p=0.011$).

The annual memory checks point in the same direction. At $\phi = 0.0005$, the conflict coefficient is negative in all four years under both models and significant in all four years. Radiation adds more explanatory power to the annual migration regressions ($\Delta R^2 = 0.010$ on average, compared with 0.006 for gravity), while gravity gives the cleaner route decomposition after origin and destination conflict are counted. Adding migration from the prior year then attenuates all conflict terms, which is why we treat the lagged specification as a conservative check based on revealed corridor choices rather than the main estimand for corridor exposure.

\clearpage

\subsection*{Spatial spillovers in conflict events}
\phantomsection
\label{si:knox_spillovers}

\refstepcounter{figure}
\paragraph*{\thefigure\ Fig.}
\label{S6_Fig}
\begin{center}
\includegraphics[width=0.95\textwidth]{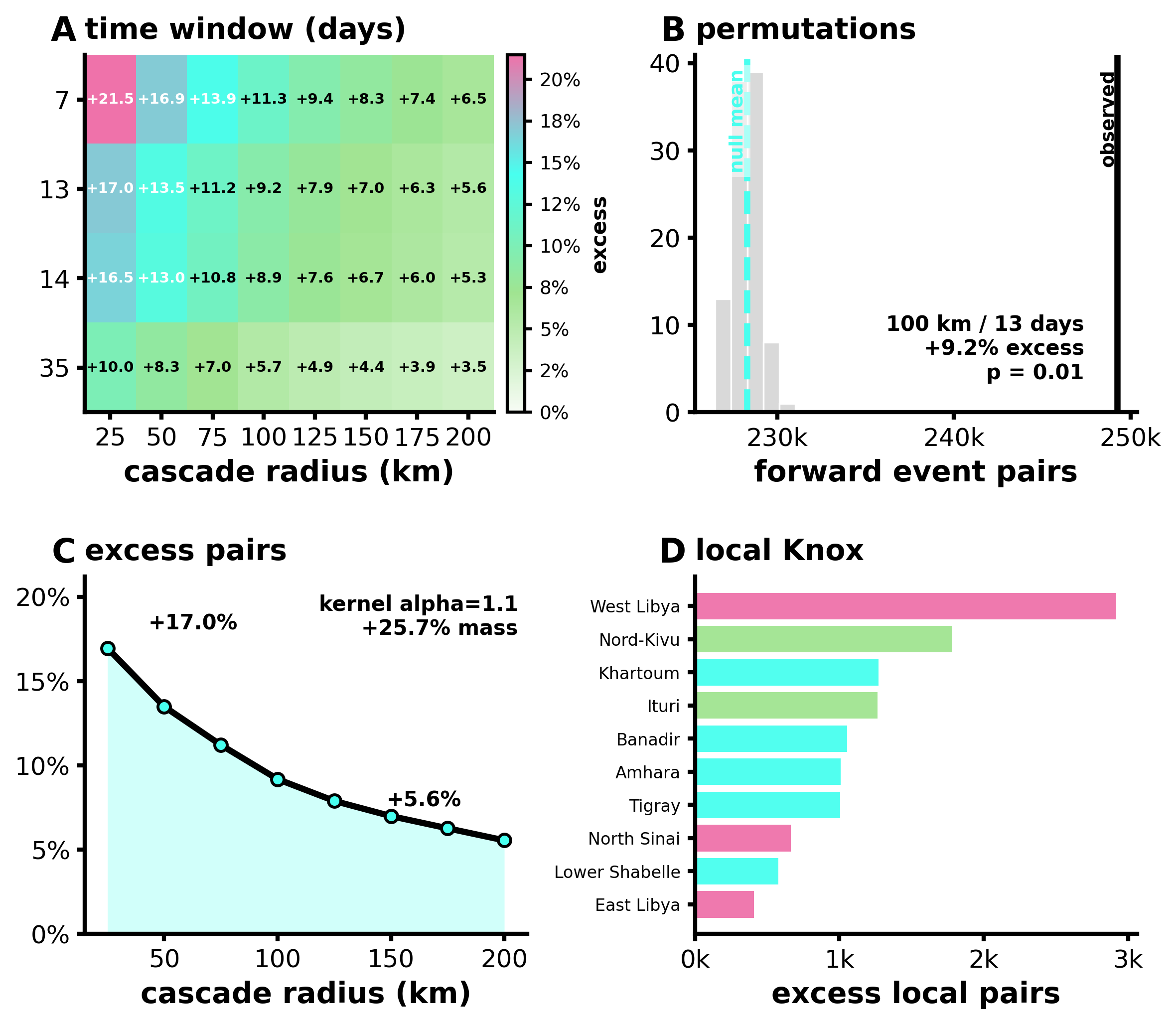}
\end{center}
\textbf{Conflict events cluster forward in both time and space.} Forward Knox tests compare observed event pairs with a null model that shuffles dates within countries and years, using only target events that occur after the seed event. A: Observed forward pairs exceed the null across the tested cascade radii and time windows derived from the Hawkes model, with the strongest excess at short distances and short windows. B: Under the main cascade window, 100\,km and 13 days, observed pairs are 9.2\% above the null ($p = 0.01$ with 99 permutations). C: Excess pairs decline with distance, from 17.0\% at 25\,km to 5.6\% at 200\,km; the kernel with a power law used in the cascade model ($\alpha = 1.1$) gives 25.7\% excess weighted mass. D: Local Knox summaries identify the places with the largest excess forward pairs, including West Libya, Nord-Kivu, Khartoum, Ituri, and Banadir. The diagnostic supports the cascade assumption that conflict risk is locally clustered after a violent event, rather than randomly redistributed within countries and years.

\clearpage

\subsection*{Population, conflict, and projected movement}
\phantomsection
\label{si:population_mu_nu}

Large cities generate more expected interaction, and cities on paths with many intervening opportunities change the radiation prediction. Population and violence may also be related. The diagnostic question is whether the conflict and mobility surfaces are simply population in another form. Using a conflict surface with long memory ($\phi = 0.002$, half-life ${\sim}347$ days), city population is only weakly related to conflict intensity and only moderately related to projected movement.

\refstepcounter{figure}
\paragraph*{\thefigure\ Fig.}
\label{S7_Fig}
\begin{center}
\includegraphics[width=0.95\textwidth]{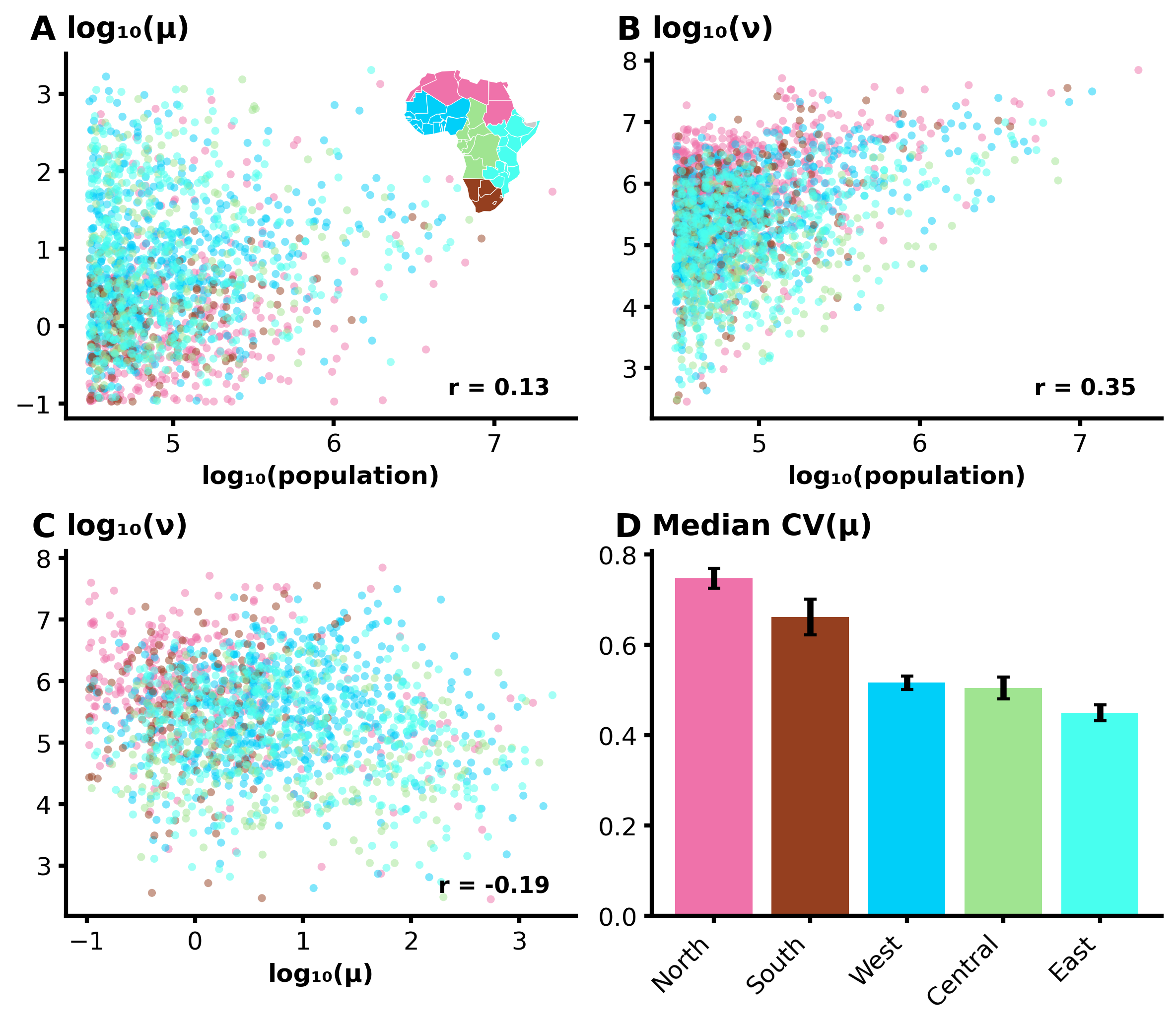}
\end{center}
\textbf{Population helps organize movement, but it does not explain corridor risk.} A: Among cities with nonzero conflict intensity in the long memory surface, population is only weakly associated with $\mu$ ($r=0.13$). B: Population is more clearly associated with $\nu$ ($r=0.35$), as expected, because projected movement is built from city size and network position. C: Conflict intensity and projected movement remain weakly separated ($r=-0.19$), so places with high risk and places with large projected flows are not the same object. D: The regional coefficient of variation in $\mu$ is highest in North and Southern Africa and lowest in East Africa, showing that temporal volatility is also geographically uneven. The plot is a population check, not a new mechanism: it shows that city size helps shape expected movement but does not absorb the conflict signal used in the corridor analysis.

\clearpage

\setcounter{table}{0}
\renewcommand{\thetable}{S\arabic{table}}

\refstepcounter{table}
\paragraph*{Table \thetable.}
\label{tab:top_city_mu}
\textbf{Cities with the highest average conflict intensity are concentrated in Somalia, conflict systems near the Sahel, and eastern DR Congo.} The table ranks cities by average daily Hawkes intensity across the 2010--2022 conflict panel used for Fig.~\ref{fig1}\textbf{A}. Peak $\mu$ captures acute shocks, while average $\mu$ separates persistent exposure from brief spikes; Mogadishu therefore sits above cities such as Juba and Seyoram, where exceptional peaks are less sustained.

\begin{center}
{\sitablefont
\begin{tabular*}{\textwidth}{@{\extracolsep{\fill}}lllrrrr@{}}
\hline
City & Country & Region & Population & Peak $\mu$ & Trough $\mu$ & Average $\mu$ \\
\hline
Mogadishu & Somalia & East & 1.7M & 763.9 & 2.9 & 192.5 \\
Benghazi & Libya & North & 594k & 486.2 & 0.0 & 60.1 \\
Maiduguri & Nigeria & West & 1.0M & 665.0 & 0.0 & 58.1 \\
Al-Arish & Egypt & North & 174k & 386.7 & 0.0 & 55.0 \\
Juba & South Sudan & East & 382k & 2046.6 & 0.0 & 39.1 \\
Rafah & Egypt & North & 78k & 307.2 & 0.0 & 34.8 \\
Surt & Libya & North & 59k & 603.4 & 0.0 & 33.8 \\
Afgooye & Somalia & East & 48k & 166.1 & 0.0 & 32.8 \\
Beni & DR Congo & Central & 327k & 203.5 & 0.0 & 32.7 \\
Merca & Somalia & East & 33k & 100.5 & 0.0 & 32.7 \\
Bangui & Central African Rep. & Central & 946k & 685.0 & 0.0 & 30.3 \\
Seyoram & Nigeria & West & 40k & 1661.6 & 0.0 & 27.7 \\
Kismaayo & Somalia & East & 104k & 127.8 & 0.0 & 27.7 \\
Baledweyne & Somalia & East & 177k & 126.9 & 0.0 & 25.3 \\
Gambaru [NGA] & Cameroon & Central & 61k & 589.8 & 0.0 & 24.9 \\
Nyala & Sudan & North & 565k & 419.1 & 0.0 & 21.5 \\
Baydhaba & Somalia & East & 170k & 71.5 & 0.0 & 19.9 \\
Galcaayo & Somalia & East & 183k & 133.1 & 0.0 & 19.0 \\
Al-Fashir & Sudan & North & 228k & 277.6 & 0.0 & 18.4 \\
Oicha & DR Congo & Central & 117k & 150.8 & 0.0 & 18.2 \\
\hline
\end{tabular*}
}
\end{center}

\refstepcounter{table}
\paragraph*{Table \thetable.}
\label{tab:mu_nu_regional}
\textbf{Conflict intensity and projected movement are only weakly coupled.} Spearman correlations between road conflict intensity $\mu$ and projected movement from radiation $\nu$ show that roads with high risk are not automatically roads that carry large flows. The relationship is positive overall but very weak, positive in Central and North Africa, not statistically clear in East Africa, negative in Southern and West Africa, and indistinguishable from zero for the small set of edges crossing countries or regions.

\begin{center}
{\sitablefont
\begin{tabular*}{\textwidth}{@{\extracolsep{\fill}}lrrr@{}}
\hline
Sample & $n$ roads & Spearman $\rho$ & $p$ value \\
\hline
All analysed roads & 5,780 & 0.066 & $<0.001$ \\
Central Africa & 581 & 0.205 & $<0.001$ \\
East Africa & 955 & -0.052 & 0.111 \\
North Africa & 2,058 & 0.045 & 0.042 \\
Southern Africa & 781 & -0.183 & $<0.001$ \\
West Africa & 1,356 & -0.276 & $<0.001$ \\
Within countries & 5,539 & 0.059 & $<0.001$ \\
Across countries & 241 & 0.001 & 0.984 \\
Across regions & 49 & -0.119 & 0.416 \\
\hline
\end{tabular*}
}
\end{center}

\refstepcounter{table}
\paragraph*{Table \thetable.}
\label{tab:conflict_migration_regional_breakdown}
\textbf{Corridor conflict is negatively associated with migration in pooled and annual models, and for routes crossing regions.} Estimates use the current radiation corridor specification across many cities with $\phi = 0.0005$. Annual coefficients are negative in every year from 2019 to 2022. The pooled estimate for routes crossing regions is also negative, while routes within the same region are not distinguishable from zero and Southern Africa is the main regional exception in this simple split. Standard errors are clustered by directed country pair. Stars mark $p < 0.001$ (***), $p < 0.01$ (**), and $p < 0.05$ (*).

\begin{center}
{\sitablefont
\begin{tabular*}{\textwidth}{@{\extracolsep{\fill}}llrrrr@{}}
\hline
Model & Subset & $n$ & $\beta_{\mathrm{conflict}}$ & SE & $R^2$ \\
\hline
Pooled & All pairs & 4,071 & -0.111*** & 0.030 & 0.283 \\
Year & 2019 & 1,017 & -0.117*** & 0.030 & 0.283 \\
Year & 2020 & 1,018 & -0.109*** & 0.029 & 0.306 \\
Year & 2021 & 1,018 & -0.107** & 0.033 & 0.278 \\
Year & 2022 & 1,018 & -0.099** & 0.033 & 0.269 \\
Origin region & North & 416 & -0.374*** & 0.091 & 0.372 \\
Origin region & West & 1,391 & -0.097* & 0.047 & 0.377 \\
Origin region & Central & 956 & -0.048 & 0.068 & 0.118 \\
Origin region & East & 1,096 & -0.014 & 0.057 & 0.345 \\
Origin region & South & 212 & 0.434* & 0.177 & 0.428 \\
Path type & Across regions & 2,288 & -0.095* & 0.041 & 0.124 \\
Path type & Same region & 1,783 & 0.021 & 0.039 & 0.469 \\
\hline
\end{tabular*}
}
\end{center}

\refstepcounter{table}
\paragraph*{Table \thetable.}
\label{tab:cascade_region_tail}
\textbf{Cascade risk is a frequency problem in West Africa and a severity problem in East and Southern Africa.} Across 10,000 cascade simulations with a power law, 80.9\% produce at least one isolated city. The full distribution has a heavy tail: $P(>1\mathrm{M}) = 18.6\%$, $P(>5\mathrm{M}) = 2.7\%$, and $P(>10\mathrm{M}) = 0.4\%$. Among nonzero cascades, West Africa contributes the largest number of events, East Africa has the highest typical impact, and Southern Africa has the widest upper tail.

\begin{center}
{\sitablefont
\begin{tabular*}{\textwidth}{@{\extracolsep{\fill}}lrrrrrr@{}}
\hline
Dominant region & Nonzero events & Share & Mean isolated (M) & Median (M) & P95 (M) & Max (M) \\
\hline
West & 3,218 & 39.8\% & 0.87 & 0.31 & 3.07 & 13.29 \\
East & 2,080 & 25.7\% & 1.11 & 0.92 & 2.37 & 7.52 \\
South & 1,831 & 22.6\% & 0.94 & 0.25 & 8.45 & 11.09 \\
Central & 508 & 6.3\% & 0.87 & 0.68 & 2.11 & 8.86 \\
North & 451 & 5.6\% & 0.15 & 0.08 & 0.37 & 1.95 \\
\hline
\end{tabular*}
}
\end{center}

\refstepcounter{table}
\paragraph*{Table \thetable.}
\label{tab:percolation_thresholds}
\textbf{A small threshold change moves the road network through the fragmentation zone.} Removing only the roads most exposed to conflict leaves the giant connected component nearly intact. The transition appears between $\mu_e > 20$ and $\mu_e > 10$: removing 874 additional roads, 9.5\% of the network, moves the mean GCC from 85.1\% to 39.2\% of nodes and sharply raises Monte Carlo variability. GCC is measured by node count, and CV reports the variation across noisy threshold runs.

\begin{center}
{\sitablefont
\begin{tabular*}{\textwidth}{@{\extracolsep{\fill}}rrrrrr@{}}
\hline
Removed if $\mu_e >$ & Roads removed & Removed roads & Mean GCC nodes & GCC share & CV \\
\hline
80 & 75 & 0.8\% & 7,310 & 99.3\% & 0.0\% \\
60 & 127 & 1.4\% & 7,277 & 98.9\% & 0.0\% \\
50 & 184 & 2.0\% & 7,245 & 98.4\% & 0.1\% \\
40 & 295 & 3.2\% & 7,142 & 97.0\% & 0.2\% \\
30 & 568 & 6.2\% & 6,896 & 93.7\% & 0.3\% \\
20 & 1,163 & 12.7\% & 6,265 & 85.1\% & 1.1\% \\
15 & 1,564 & 17.1\% & 5,200 & 70.6\% & 16.2\% \\
10 & 2,037 & 22.2\% & 2,888 & 39.2\% & 22.0\% \\
5 & 2,937 & 32.1\% & 947 & 12.9\% & 19.7\% \\
\hline
\end{tabular*}
}
\end{center}

\refstepcounter{table}
\paragraph*{Table \thetable.}
\label{tab:top_vulnerable_corridors}
\textbf{Examples of vulnerable road segments combine conflict exposure, projected movement, and weak substitutes.} Rows are selected from roads with nonzero projected radiation movement and $\mu_e \geq 10$, using a vulnerability score that ranks high conflict intensity, high projected movement, and long or missing detours. Finite detours are the added shortest-path distance after removing the road; ``no substitute'' means the endpoints disconnect in the road network. The table is illustrative rather than exhaustive, but it makes the policy margin explicit: vulnerability is highest where exposure and sparse redundancy overlap.

\begin{center}
{\sitablefont
\begin{tabular*}{\textwidth}{@{\extracolsep{\fill}}llrrll@{}}
\hline
Road segment & Region & $\mu_e$ & $\nu_e$ & Substitute & Network relevance \\
\hline
Oicha--road junction & Central & 126.0 & 0.0149 & +1{,}147 km & high-risk road with long finite detour \\
Baraka--Uvira & Central & 121.9 & 0.0113 & +2{,}363 km & city-to-city link with extreme detour \\
Gao--Mopti-Sevare & West & 69.7 & 0.0101 & +770 km & Sahel segment in fragmentation zone \\
Bukavu--Cyangugu & Cross-region & 69.4 & 0.0129 & +368 km & cross-border Great Lakes connector \\
Gao--Bourem & West & 33.5 & 0.0192 & +4{,}232 km & high-flow road with extreme detour \\
Kismaayo--Merca & East & 218.0 & 0.0045 & +2{,}961 km & high-conflict Horn corridor segment \\
Afgooye--Merca & East & 449.4 & 0.0044 & +3{,}601 km & intense exposure with long substitute \\
Manzini--road junction & South & 28.6 & 0.0048 & no substitute & lower-risk but nonredundant connector \\
\hline
\end{tabular*}
}
\end{center}

\clearpage

\subsection*{Annual fragmentation and detours}
\phantomsection
\label{si:annual_percolation_detours}

\refstepcounter{figure}
\paragraph*{\thefigure\ Fig.}
\label{S8_Fig}
\begin{center}
\includegraphics[width=0.98\textwidth]{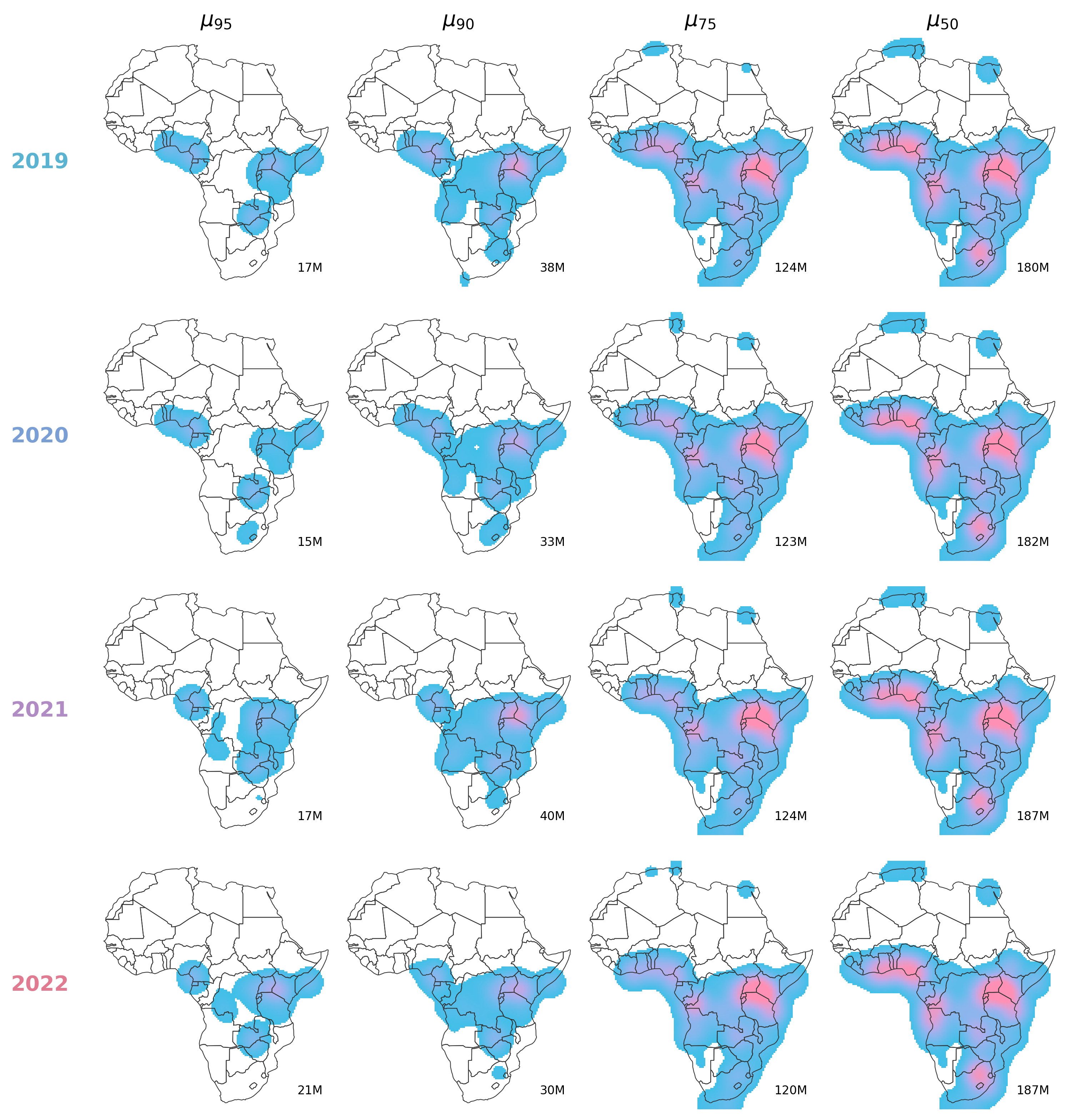}
\end{center}
\textbf{Annual percolation maps show recurring vulnerability in sparse corridors.} Rows use yearly Hawkes risk surfaces from 2019 to 2022, and columns remove roads above pooled risk percentiles, from the most extreme 5\% of roads ($\mu_{95}$) to the upper half of the risk distribution ($\mu_{50}$). At the highest threshold, isolated population remains limited, between 15M and 21M across years. By $\mu_{75}$ it rises to roughly 120M--124M, and by $\mu_{50}$ it reaches 180M--187M. The affected geography changes with the annual conflict surface, but the same sparse corridors recur, especially around East Africa, the Sahel, and parts of Southern Africa.

\clearpage

\refstepcounter{figure}
\paragraph*{\thefigure\ Fig.}
\label{S9_Fig}
\begin{center}
\includegraphics[width=0.98\textwidth]{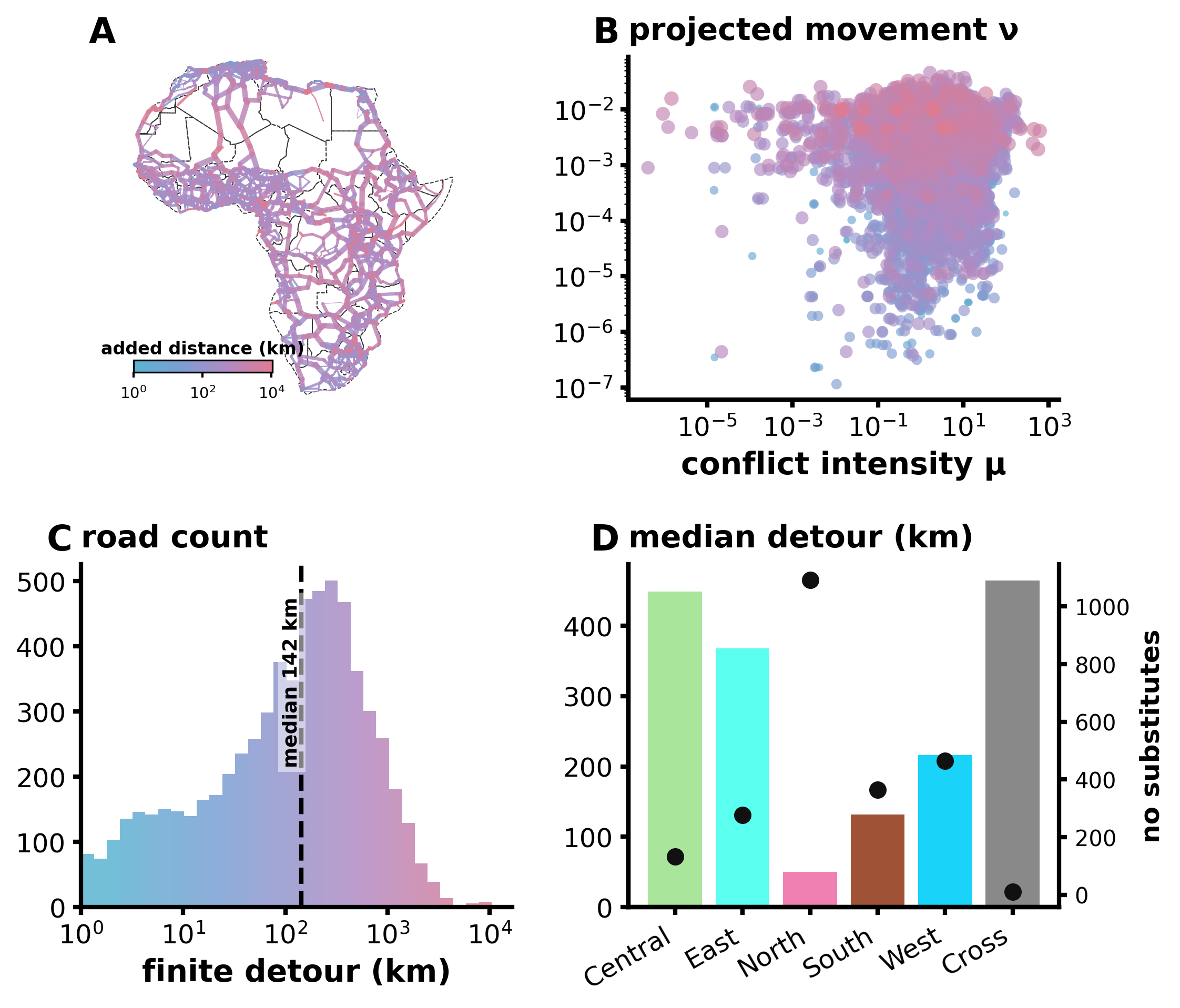}
\end{center}
\textbf{Rerouting creates risk even when roads remain connected.} A: Road width shows projected movement from radiation $\nu$; color shows the added distance if the road is removed, with roads that have no substitute assigned the maximum color value. B: Roads with high conflict and large flows are not always the largest detour risks, so vulnerability depends on redundancy as well as exposure. C: Where substitutes exist, removal adds a median 142 km and a mean 323 km; even the median detour is a substantial added journey. D: Central, East, and roads crossing regions have the longest median finite detours, while roads with no substitute are most numerous in North Africa (1,092 roads). The figure shows that road loss related to conflict can matter through prolonged rerouting as well as network fragmentation.

\clearpage

\subsection*{News reports of corridor disruption}
\phantomsection
\label{si:gdelt_disruption}

\refstepcounter{figure}
\paragraph*{\thefigure\ Fig.}
\label{S10_Fig}
\begin{center}
\includegraphics[width=0.95\textwidth]{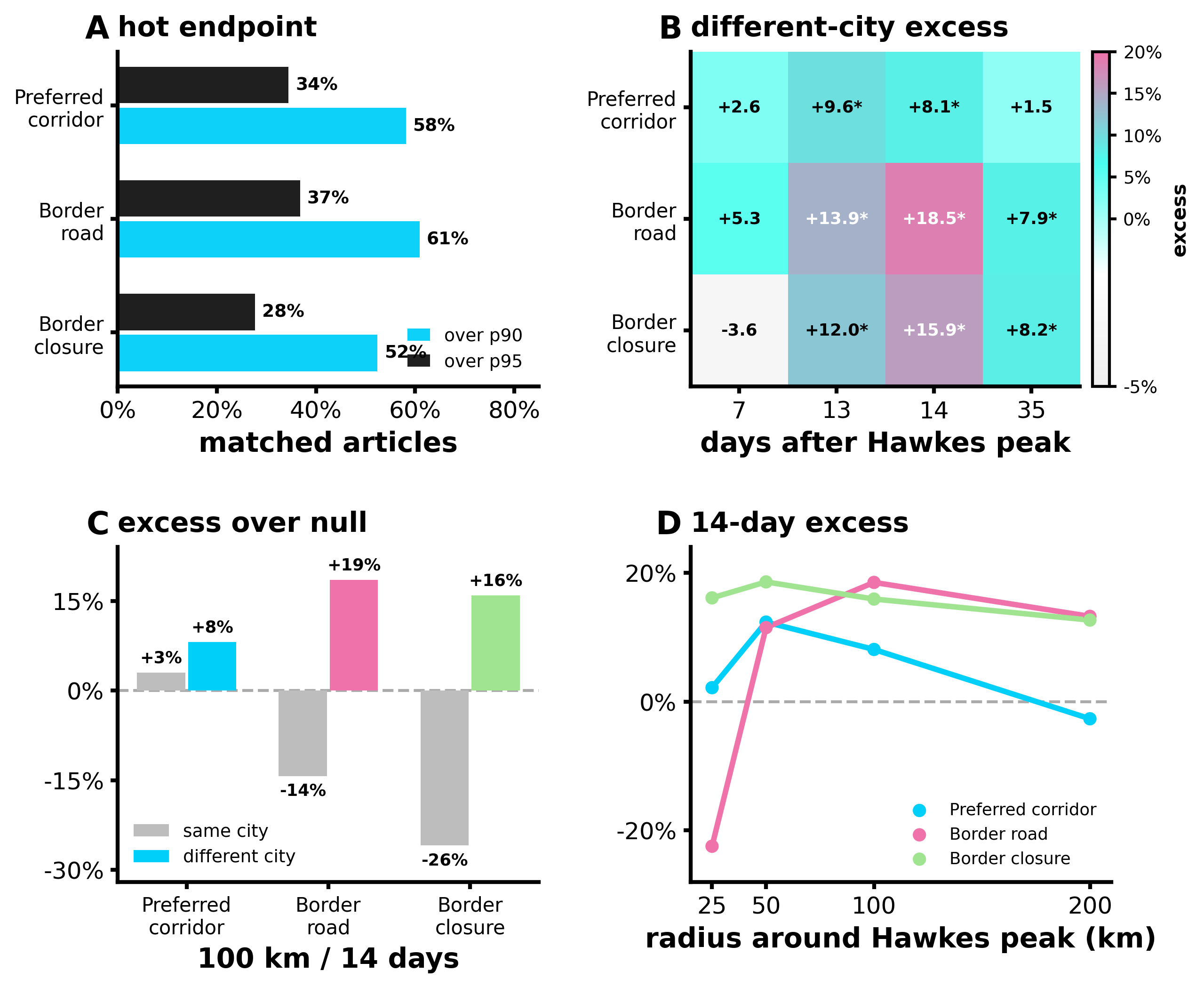}
\end{center}
\textbf{Reports of disruption spread beyond the conflict city.} A: News reports extracted from the GDELT Project that mention corridor, border-road, or border-closure disruption frequently name at least one city with high Hawkes conflict intensity. B: In a forward bivariate Knox test, peaks in city-level Hawkes intensity are followed by excess disruption mentions in different nearby cities over the next 13--14 days, compared with a country-month date-shuffled null. C: The excess is not driven by same-city continuation: at 100\,km and 14 days, same-city pairs are weak or negative for the border categories, while different-city pairs are positive. D: The 14-day different-city excess is strongest around 50--200\,km, consistent with corridor and border disruption appearing in the surrounding city system rather than only at the conflict seed. The figure is a reporting-based robustness check rather than a primary outcome model; GDELT article-city mentions are noisy and clustered by source, but they provide independent evidence that conflict heat and mobility disruption language move together in space and time.

}

\clearpage

\section*{Acknowledgments}
 

\bibliography{references} 

\end{document}